\documentclass[12pt]{article}
\usepackage[english]{babel}
\usepackage{natbib}
\usepackage{comment}
\usepackage{float}
\usepackage[hidelinks]{hyperref}
\usepackage{mathrsfs}
\usepackage{enumitem}
\usepackage[font={small,it}]{caption}
\usepackage{amsmath,amsfonts,amsthm,amssymb}
\usepackage{bm,rotating,multirow,dsfont,graphicx}
\usepackage[usenames, dvipsnames]{color}
\usepackage{url}
\usepackage{multicol}
\usepackage{multirow}
\usepackage[T1]{fontenc}
\usepackage{flafter}
\usepackage{appendix}
\usepackage{subfigure}
\usepackage{xcolor}
\usepackage{soul}
\usepackage{setspace}
\usepackage{booktabs}
\usepackage{algorithm}
\usepackage{algpseudocode}
\usepackage{subcaption}
\usepackage{mathtools}
\usepackage{cancel}
\usepackage{graphicx}
\usepackage{svg}
\makeatletter
\def\hlinewd#1{%
	\noalign{\ifnum0=`}\fi\hrule \@height #1 %
	\futurelet\reserved@a\@xhline}
\makeatother
\def\spacingset#1{\renewcommand{\baselinestretch}{#1}\small\normalsize}\spacingset{1}
\def\@roman#1{\romannumeral #1}

\begin{document}

\title{Latent space models for networks with nodal multiplicative effects}

\date{2026}

\author{
Carlos Nosa\footnote{Corresponding author: cnosa@unal.edu.co.}\qquad \qquad 
Juan Sosa \\
Facultad de Ciencias\\
Universidad Nacional de Colombia, sede Bogotá, Colombia
}

\maketitle

\begin{abstract}
Latent space models represent network nodes as points in a geometric space, with connection probabilities determined by distances between latent positions under a fixed metric, typically Euclidean, spherical, or hyperbolic. We generalize the classical formulation by introducing nodal multiplicative effects motivated by a local deformation of the latent metric. This modification approximates a conformal deformation of the metric tensor while preserving the logistic predictor and the geometric interpretability of the model, thereby capturing additional structural heterogeneity without altering the global reference geometry. We study the generative behavior of the proposed model through simulation experiments and develop an optimization-based inference scheme derived from a hierarchical Bayesian formulation with parameter regularization. Applications to eight real networks show that the proposed approach increases the generative flexibility of classical latent space models and more accurately reproduces several topological properties observed in real networks.
\end{abstract}

\spacingset{1.1} 

\newpage

\section{Introduction}

The study of complex systems through the interactions among their components is a central focus in many scientific disciplines, including epidemiology, sociology, systems biology, economics, and data science \citep{newman2018networks, barabasi2016network}. Such systems are commonly represented as networks or graphs, where nodes correspond to entities and edges describe the relationships between them. Statistical network analysis aims, among other objectives, to characterize structural patterns, understand link formation mechanisms through probabilistic models, and predict unobserved or future connections based on structural properties \citep{goldenberg2010survey, kolaczyk2009statistical}.

A widely studied approach for addressing these dependencies is the class of latent space models, which propose representing the network in an unobserved geometric space of reduced dimension \citep{hoff_latent_2002}. Inspired by the tradition of latent variable models used to simplify complex dependency structures, these approaches assume that, conditional on the latent positions, the edges are independent. That is, the observed dependency in the network is induced through the geometric proximity between nodes in this latent space.

In this approach, each node is associated with an unobserved position in a low-dimensional metric space, commonly interpreted as a “social space.” The probability of a connection between two nodes is a decreasing function of the distance between their latent positions. This construction provides an intuitive geometric representation of the network and allows it to be visualized as a map of latent relationships. In particular, phenomena such as homophily (the tendency of similar individuals to connect) naturally emerge as spatial proximity, while transitivity derives from the triangular inequality of the underlying metric \citep{hoff_latent_2002, hoff2008}.

In addition to their interpretability, latent space models offer a parsimonious parameterization of structural dependency: rather than explicitly modeling combinatorial configurations of edges, the dependency is encapsulated in the geometry of the latent space. This property has given rise to an extensive body of literature that includes extensions for community structures, temporal dynamics, multilayer networks, and other types of non-standard relational data \citep{goldenberg2010survey, sewell2015latent}.

Traditionally, most of these models have used Euclidean spaces (such as $\mathbb{R}^d$ with $d$ a positive integer) because of their mathematical familiarity and computational ease \citep{hoff_latent_2002}. However, subsequent work has shown that the choice of underlying geometry can significantly influence the model’s ability to reproduce global properties observed in real networks \citep{krioukov2010hyperbolic, smith2019,nosa2026spherical}. In particular, hyperbolic spaces allow for capturing degree heterogeneity and hierarchical organization, whereas other geometries may impose different structural constraints. Consequently, the geometry of the latent space is not merely a technical tool, but a structural component that determines the dependency regime induced in the observed network.

In the classical formulation, the latent space is modeled as a manifold with a constant metric, typically Euclidean, spherical, or hyperbolic \citep{krioukov2010hyperbolic, smith2019}. The choice of geometry determines global structural properties of the generated network: Euclidean geometry favors approximately homogeneous structures \citep{hoff_latent_2002}, hyperbolic geometry allows for modeling degree heterogeneity and hierarchical structures \citep{krioukov2010hyperbolic}, while spherical geometry imposes global constraints of positive curvature \citep{smith2019, nosa2026spherical}. However, in all these cases, the underlying metric is homogeneous in space, which implies that the “geometric cost” of displacement is identical in all regions of the latent space. This assumption of homogeneity can be restrictive when modeling networks with pronounced local structural heterogeneity; nodes with atypical behavior or regions of latent space with varying connection densities cannot be represented without modifying the global geometry or introducing additional effects, as in bilinear or multiplicative models \citep{hoff_bilinear_2005, hoff2009multiplicative}.

In this paper, we propose a generalization of latent space models by incorporating node-specific multiplicative effects into the distance function, which are motivated by a local metric deformation. Specifically, we introduce a node-dependent multiplicative term into the distance function that determines the logistic predictor. This modification can be interpreted as an approximation to a conformal deformation of the metric tensor of the ambient geometry \citep{lee2018riemannian}. In this way, we preserve the global geometric structure (Euclidean, spherical, or hyperbolic), while allowing for local variations in the metric scale induced by each node. Thus, the resulting model preserves the geometric interpretability of the classical approach \citep{hoff_latent_2002} and extends its expressive power by allowing specific nodes to effectively expand or contract their latent neighborhood. This induces more heterogeneous connectivity patterns without altering the global geometry of the space.

The proposed nodal metric effects differ fundamentally from the additive nodal random effects and degree-correction mechanisms commonly used in network models \citep{hoff_bilinear_2005, karrer2011stochastic}. Whereas additive effects modify the baseline propensity of a node to form ties by introducing node-specific terms outside the distance function, our approach directly modifies the latent distance itself through a local metric deformation. Consequently, two nodes with similar latent positions may still exhibit different connectivity patterns because the effective distance between them depends on their local metric scales. The proposed model therefore captures node-level heterogeneity through the geometry of the latent space rather than through additive popularity effects.

We develop a hierarchical Bayesian formulation of the model. In this framework, the latent geometry is treated as a fixed model choice, while the latent positions and nodal effects are estimated following the conceptual framework of \cite{hoff2011hierarchical, sewell2015latent}. The resulting objective function is optimized using manifold optimization techniques \citep{smith2014optimization, absil2008optimization} to obtain point estimates. We also analyze the generative behavior of the model under different configurations and compare its ability to reproduce relevant topological statistics against classical models, closely following the methodology of \cite{smith2019}.

The main contributions of this work are the development of a geometrically motivated extension of latent space models based on interpretable nodal multiplicative effects, which approximate a conformal deformation of the latent geometry, as well as the establishment of conditions for model identifiability and estimability under different ambient geometries. We also develop a manifold optimization scheme whose objective function is derived from the posterior log-density under a hierarchical Bayesian formulation. Through simulations and comparative studies, we show that the proposed model increases generative flexibility and improves the reproduction of structural properties observed in real networks. Finally, we compare the proposed and classical models using eight real-world networks and evaluate their goodness of fit and performance across several relevant criteria.

The rest of the article is organized as follows. In Section~\ref{sec2}, we present the geometric and statistical foundations of the model. In Section~\ref{sec3}, we formally introduce the proposed new statistical model for networks and discuss aspects of identifiability. In Section~\ref{sec4}, we develop the inference methods. In Section~\ref{sec5}, we illustrate the model’s performance through empirical applications. Finally, in Section~\ref{sec6}, we discuss the conclusions of this work and possible extensions.

\section{Geometric and statistical fundamentals} \label{sec2}

Latent space models for networks are based on the idea of representing each node in a network as a point on a manifold equipped with a metric \citep{smith2019, sosa_review_2021}. In this framework, the probability of a connection between two nodes depends on the geodesic distance induced by that metric. This formulation allows us to interpret structural properties of the network in geometric terms, where greater similarity in latent space translates to a higher probability of a link \citep{hoff_latent_2002}.

Let us first consider the geometric framework of latent spaces in the models mentioned above; the geometric concepts and definitions presented in this section are based on \cite{lee2018riemannian}. Let $(\mathbb{M}_{\kappa}^{d}, g_\kappa)$ be a complete and connected manifold of constant curvature $\kappa \in \mathbb{R}$ and effective dimension $d$, where $\mathbb{M}_\kappa^d \subset \mathbb{R}^{d+1}$ is the subset that defines the manifold as a submanifold embedded in $\mathbb{R}^{d+1}$, and $g_\kappa$ is the metric tensor such that, for each $\mathbf{x}\in\mathbb{M}_\kappa^{d}$, $g_\kappa$ assigns a function $g_\kappa[\mathbf{x}]: T_\mathbf{x}\mathbb{M}_\kappa^d\times T_{\mathbf{x}}\mathbb{M}_\kappa^d \to \mathbb{R}$ that is bilinear, symmetric, and positive definite, where $T_{\mathbf{x}}\mathbb{M}_\kappa^d$ is the tangent space to the manifold $\mathbb{M}_\kappa^d$ at the point $\mathbf{x}$. We denote by $d_{\kappa}(\cdot,\cdot)$ the geodesic distance induced by the metric tensor $g_\kappa$. In general, given a manifold $(\mathbb{M}_\kappa^d,g_\kappa)$, the geodesic distance between two points $\mathbf{x},\mathbf{y}\in\mathbb{M}_\kappa^d$ is defined as:
\begin{equation*}
d_{\kappa}(\mathbf{x},\mathbf{y})
=
\underset{\gamma \in \Gamma}{\inf}
\int_0^1 
\sqrt{
g_\kappa[\gamma(t)]\big(\dot{\gamma}(t),\dot{\gamma}(t)\big)
}
\, dt, 
\end{equation*}

where $\Gamma = \{\gamma \in \mathcal{C}^1([0,1],\mathbb{M}_\kappa^d):\;
\gamma(0)=\mathbf{x},\;
\gamma(1)=\mathbf{y}\}$ is the set of all continuously differentiable curves defined on the interval $[0,1]$ such that their image starts at $\mathbf{x}$ and ends at $\mathbf{y}$. When the manifold is complete, this infimum is attained by a minimizer geodesic \citep{lee2018riemannian}.

\begin{figure}[!htb]
\centering
\includegraphics[width=0.75\linewidth]{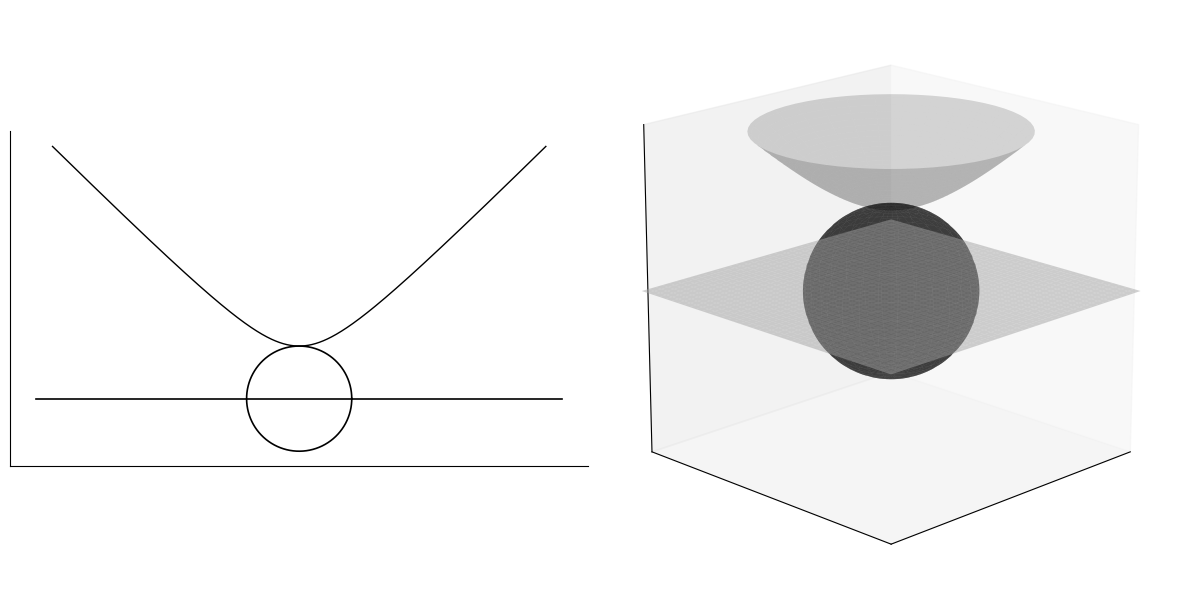}
\caption{Geometric representation of constant-curvature manifolds embedded in Euclidean spaces: on the left, $\mathbb{R}^1$ (line), $\mathbb S^1$ (circle), and $\mathbb H^1$ (hyperboloid) embedded in $\mathbb{R}^2$; on the right, $\mathbb{R}^2$ (plane), $\mathbb S^2$ (spherical shell), and $\mathbb H^2$ (hyperboloid) embedded in $\mathbb{R}^3$.}
\label{fig:latent_spaces_dimensions}
\end{figure}

In classical latent space models with constant curvature, the metric tensor $g_\kappa$ is homogeneous in the sense that its expression does not depend explicitly on the point $\mathbf{x}\in\mathbb{M}_\kappa^d$, but only on the global geometry determined by $\kappa$. Consequently, the local structure of the space is identical at all points; that is, the “infinitesimal cost” of displacement is the same in any region of the manifold. From a statistical perspective, this homogeneity implies that the local connectivity properties in the latent model are determined exclusively by the relative positions of the nodes (that is, by the geodesic distance $d_{\kappa}(\mathbf{z}_i,\mathbf{z}_j)$) and not by local variations in the metric. The global geometry (or that of the ambient space) controls the macroscopic structure induced in the direct network model, whereas individual heterogeneity can only be introduced through displacements in the latent space and not through local metric deformations. To illustrate the concepts mentioned above, well-known cases of manifolds are described below, distinguishing them according to the sign of the curvature. A visual representation is provided in Figure \ref{fig:latent_spaces_dimensions}, which shows representations of Euclidean, spherical, and hyperbolic manifolds of effective dimensions one and two, embedded in a higher-dimensional Euclidean space.

\begin{itemize}
\item \textbf{Euclidean case ($\kappa = 0$).} If $\kappa = 0$, the manifold $\mathbb{M}_0^d$ is isometric to $\mathbb{R}^d$, which is a complete but non-compact space. The manifold is equipped with the standard Euclidean metric given by the tensor:
\begin{equation*}
g_{0}[\mathbf{x}](\mathbf{u},\mathbf{v}) 
= \langle \mathbf{u},\mathbf{v} \rangle,
\quad 
\mathbf{u},\mathbf{v} \in T_\mathbf{x}\mathbb{R}^d \cong \mathbb{R}^d.
\end{equation*}

Using the metric tensor, the length of a differentiable curve $\gamma:[0,1]\to\mathbb{R}^d$ is defined as $L_{0}(\gamma) = \int_0^1 \sqrt{ g_0[\gamma(t)]\big(\dot{\gamma}(t),\dot{\gamma}(t)\big) } \, dt$. Since the metric tensor coincides with the usual inner product, we obtain $g_0[\gamma(t)](\dot{\gamma}(t),\dot{\gamma}(t)) = \langle \dot{\gamma}(t),\dot{\gamma}(t)\rangle = \|\dot{\gamma}(t)\|^2$, and therefore:
\begin{equation*}
L_{0}(\gamma) = \int_0^1 \|\dot{\gamma}(t)\| \, dt,
\end{equation*}
which is the classical Euclidean length of the curve. The geodesic distance between $\mathbf{x}$ and $\mathbf{y}$ is defined as the minimum value of $L_{0}(\gamma)$ among all curves connecting these two points \citep{colley1998vector}. In Euclidean space, the minimizing curve is the line segment $\gamma(t) = (1-t)\mathbf{x} + t\mathbf{y}$, for which $\dot{\gamma}(t)=\mathbf{y}-\mathbf{x}$ is constant. Consequently, $L_{0}(\gamma) = \int_0^1 \|\mathbf{y}-\mathbf{x}\| \, dt = \|\mathbf{y}-\mathbf{x}\|$. Therefore, the geodesic distance induced by the Euclidean metric tensor coincides with the Euclidean norm $d_0(\mathbf{x},\mathbf{y}) = \|\mathbf{x}-\mathbf{y}\|$.

To show that the straight line is indeed the curve that minimizes the
distance between $\mathbf{x}$ and $\mathbf{y}$, let $\gamma:[0,1]\to\mathbb{R}^d$
be any differentiable curve such that $\gamma(0)=\mathbf{x}$ and $\gamma(1)=\mathbf{y}$. Then $\mathbf{y}-\mathbf{x}= \gamma(1)-\gamma(0)= \int_0^1 \dot{\gamma}(t)\,dt$. Taking the Euclidean norm and using the triangle inequality for integrals,
\begin{equation*}
\|\mathbf{y}-\mathbf{x}\|
= \left\|\int_0^1 \dot{\gamma}(t)\,dt\right\|
\leq \int_0^1 \|\dot{\gamma}(t)\|\,dt
= L_{0}(\gamma).
\end{equation*}
Therefore, the length of any curve connecting $\mathbf{x}$ to $\mathbf{y}$ satisfies $L_{0}(\gamma) \geq \|\mathbf{y}-\mathbf{x}\|$, and since the line segment between $\mathbf{x}$ and $\mathbf{y}$ has length $\|\mathbf{y}-\mathbf{x}\|$, it is the minimizing curve.

\item \textbf{Spherical case ($\kappa > 0$).}
If $\kappa > 0$, let $R = 1/\sqrt{\kappa}$. The model space is the sphere of radius $R$:
\begin{equation*}
\mathbb{M}_\kappa^d
=
\left\{ \mathbf{x} \in \mathbb{R}^{d+1} : 
\langle \mathbf{x},\mathbf{x} \rangle = R^2 \right\},
\end{equation*}
equipped with the metric induced by the Euclidean inner product of the ambient space $\mathbb{R}^{d+1}$. The metric tensor is defined as the restriction of the inner product to the tangent space:
\begin{equation*}
g_{\kappa}[\mathbf{x}](\mathbf{u},\mathbf{v}) 
= 
\langle \mathbf{u},\mathbf{v} \rangle,
\text{ where }
\mathbf{u},\mathbf{v} \in T_{\mathbf{x}}\mathbb{M}_\kappa^d 
\text{ for all } \mathbf{x}\in \mathbb{M}_\kappa^d.
\end{equation*}
The geodesic distance is given by:
\begin{equation*}
d_\kappa(\mathbf{x},\mathbf{y})
=
R \, 
\operatorname{arccos}\!\left(
\frac{\langle \mathbf{x},\mathbf{y} \rangle}{R^2}
\right),
\end{equation*}
and the space is compact, complete, and has a constant positivecurvature.

\item \textbf{Hyperbolic case ($\kappa < 0$).}
If $\kappa < 0$, let $R = 1/\sqrt{-\kappa}$. The model space can be represented by a single-sheet hyperboloid:
\begin{equation*}
\mathbb{M}_\kappa^d
=
\left\{
\mathbf{x} \in \mathbb{R}^{d+1} :
\langle \mathbf{x},\mathbf{x} \rangle_L = -R^2,
\; x_{d+1} > 0
\right\}, 
\end{equation*}
where:
\begin{equation*}
\langle \mathbf{x},\mathbf{y} \rangle_L
=
\sum_{i=1}^{d} x_i y_i - x_{d+1} y_{d+1}
\end{equation*}
is the Lorentz inner product. The metric tensor is obtained by restricting the Lorentz product to the tangent space:
\begin{equation*}
g_{\kappa}[\mathbf{x}](\mathbf{u},\mathbf{v})
=
\langle \mathbf{u},\mathbf{v} \rangle_L,
\text{ where }
\mathbf{u},\mathbf{v} \in T_{\mathbf{x}}\mathbb{M}_\kappa^d \text{ for all } \mathbf{x}\in \mathbb{M}_\kappa^d.
\end{equation*}
The geodesic distance is given by:
\begin{equation*}
d_\kappa(\mathbf{x},\mathbf{y})
=
R \,
\operatorname{arccosh}
\!\left(
-\frac{\langle \mathbf{x},\mathbf{y} \rangle_L}{R^2}
\right).
\end{equation*}
This space is complete, not compact, and has constant negative curvature. For the hyperbolic geometry, we chose the single-sheet hyperboloid to symbolize the space because of its parsimonious definition and ease of computational representation; other representations can be found in \cite{ratcliffe2006foundations}.
\end{itemize}

Once the geometric foundations have been established, we move on to the statistical foundations. Indeed, let $\mathbf{Y} = (y_{i,j})$ be the adjacency matrix of an undirected binary network with $n$ nodes, where $y_{i,j} \in \{0,1\}$, $y_{i,j}=y_{j,i}$, and $y_{i,i}=0$. A statistical model for networks consists of a family of distributions $p(\mathbf{Y} \mid \boldsymbol{\theta})$ where $\boldsymbol{\theta} \in \Theta$, which describes the mechanism generating the links based on the parameters and allows us to quantify the uncertainty associated with unknown parameters \citep{kolaczyk_statistical_2020}.

In the case of an undirected binary network, under conditional independence given a latent structure, the following is assumed:
\begin{equation*}
y_{i,j} \mid s_{i,j} \overset{\text{ind}}{\sim} \operatorname{Bernoulli}\!\big(p_{i,j} = \operatorname{expit}(s_{i,j})\big),
\qquad 1 \le i < j \le n,
\end{equation*}
where $s_{i,j}$ is a predictor and the function $\operatorname{expit}(x) = \frac{1}{1+e^{-x}}$ with $x\in\mathbb{R}$ is the inverse of $\operatorname{logit}(p) = \log\left (\frac{p}{1-p}\right)$ for $p\in(0,1)$, which is the canonical link function for a generalized linear model with a Bernoulli response \citep{myers1997tutorial, mccullagh1989generalized}. In classical distance-based models, the predictor takes the form:
\begin{equation*}
s_{i,j} = \alpha_0 + \alpha_1 \mathbf{w}_{i,j} - d_{\kappa}(\mathbf{z}_i,\mathbf{z}_j),
\end{equation*}
where $\alpha_0$ controls the overall density of the network, $\alpha_1$ is a parameter that controls the effect of covariates $\mathbf{w}_{i,j}$, $d_{\kappa}$ is the geodesic distance induced from the metric tensor ${g_\kappa}$, and $\mathbf{z}_i,\mathbf{z}_j \in \mathbb{M}_\kappa^d \subset \mathbb{R}^{d+1}$ are the latent positions associated with nodes $i$ and $j$, respectively \citep{sosa_review_2021}. This specification induces a logistic decay of the connection probability as a function of the geodesic separation of the positions in latent space. In this work, we will consider only the predictor without covariates in order to study the simplest case; this model can be interpreted as a purely geometric baseline model in which the connection probability depends exclusively on proximity in latent space. In this scenario, the predictor reduces to $s_{i,j} = \alpha_0 - d_{\kappa}(\mathbf{z}_i,\mathbf{z}_j)$.

This simplification is useful from both a theoretical and computational standpoint. On the one hand, it allows us to directly study how the curvature of the latent space and the arrangement of the positions $\mathbf{z}_i$ determine global properties such as density, transitivity, and degree distribution. A possible future direction for this work is to include nodal or relational information in the predictor $s_{i,j}$.

In order to introduce local geometric heterogeneity, we propose considering a conformal deformation of the base metric tensor. Let $\xi : \mathbb{M}_\kappa^d \to (0,\infty) \subset \mathbb{R}$ be a smooth function; we define the new metric tensor as $g_\xi[\mathbf{x}] = \xi(\mathbf{x})^{2}\, g_\kappa[\mathbf{x}]$. The geodesic distance induced by this metric is:
\begin{align*}
d_{\xi}(\mathbf{x},\mathbf{y}) &= \underset{\gamma\in\Gamma}{\inf}
\int_0^1 
\sqrt{
g_\xi[\gamma(t)]\big(\dot{\gamma}(t),\dot{\gamma}(t)\big)
}
\, dt
= \underset{\gamma\in\Gamma}{\inf}
\int_0^1 
\xi(\gamma(t))\sqrt{
g_\kappa[\gamma(t)]\big(\dot{\gamma}(t),\dot{\gamma}(t)\big)
}
\, dt,
\end{align*}
that is, we calculate the geodesic distance by taking the infimum of continuously differentiable curves connecting $\mathbf{x}$ and $\mathbf{y}$. 

As can be seen, the length functional 
$$
\int_0^1 \sqrt{ g_\xi[\gamma(t)]\big(\dot{\gamma}(t),\dot{\gamma}(t)\big) } \,dt
$$ 
of the geodesic distance preserves the term of the integrand from the base geometry 
$$
\sqrt{
g_\kappa [\gamma(t)]\big(\dot{\gamma}(t),\dot{\gamma}(t)\big)
}
$$
and undergoes a multiplicative modification given by ${\xi(\mathbf{x})}$. Since this distance generally does not admit a closed-form expression, an approximation consisting of two conceptual steps is adopted. First, the geodesic of the ambient space $(\mathbb{M}_\kappa^d, g_\kappa)$ is fixed, and second, the conformal factor is linearly approximated along that trajectory. 
Indeed, let $\gamma^\ast$ be the minimizing geodesic curve between $\mathbf{x}$ and $\mathbf{y}$ in the geometry of the ambient space, that~is:
\begin{equation*}
\gamma^\ast = \arg\min_{\gamma \in \Gamma} \int_0^1 \sqrt{g_\kappa[\gamma(t)]\big(\dot{\gamma}(t), \dot{\gamma}(t)\big)}\, dt,
\end{equation*}
so that:
\begin{equation*}
d_\kappa(\mathbf{x}, \mathbf{y}) = \int_0^1 \sqrt{g_\kappa[\gamma^\ast(t)]\big(\dot{\gamma}^\ast(t), \dot{\gamma}^\ast(t)\big)}\, dt.
\end{equation*}

Under the metric defined by $g_\xi = \xi^2 g_\kappa$, the length of a curve $\gamma$ is given by:
\begin{equation*}
L_\xi(\gamma) = \int_0^1 \xi(\gamma(t)) \, \sqrt{g_\kappa[\gamma(t)]\big(\dot{\gamma}(t), \dot{\gamma}(t)\big)}\, dt.
\end{equation*}

The first step of the approach consists of restricting the variational problem to the geodesic $\gamma^\ast$ of the ambient space, that is:
\begin{equation*}
d_\xi(\mathbf{x}, \mathbf{y}) = \inf_{\gamma\in\Gamma} L_\xi(\gamma) \leq L_\xi(\gamma^\ast)=\int_0^1 \xi(\gamma^\ast(t)) \, \sqrt{g_\kappa[\gamma^\ast(t)]\big(\dot{\gamma}^\ast(t), \dot{\gamma}^\ast(t)\big)}\, dt.
\end{equation*}

Note that the term $\sqrt{g_\kappa[\gamma^\ast(t)]\big(\dot{\gamma}^\ast(t), \dot{\gamma}^\ast(t)\big)}$ corresponds to the infinitesimal length element of the base geodesic $\gamma^\ast$. If $\gamma^\ast$ is parameterized proportionally to the arc length, this term is constant and equal to $d_\kappa(\mathbf{x}, \mathbf{y})$. Consequently, it can be factored to yield:
\begin{equation*}
d_\xi(\mathbf{x}, \mathbf{y}) \leq d_\kappa(\mathbf{x}, \mathbf{y}) \int_0^1 \xi(\gamma^\ast(t))\, dt.
\end{equation*}

The second step consists of approximating the integral of the conformal factor along the trajectory using a first-order quadrature rule \citep{burden2011numerical}. Using a trapezoidal rule approximation:
\begin{equation*}
\int_0^1 \xi(\gamma^\ast(t))\, dt \approx \frac{\xi(\gamma^\ast(0)) + \xi(\gamma^\ast(1))}{2}
= \frac{\xi(\mathbf{x}) + \xi(\mathbf{y})}{2}.
\end{equation*}

Substituting this approximation into the previous expression yields:
\begin{equation*}
d_{\xi}(\mathbf{x},\mathbf{y}) 
\approx \tilde{d}_{\xi}(\mathbf{x},\mathbf{y}) :=
\frac{{\xi(\mathbf{x})} + {\xi(\mathbf{y})}}{2}
\, d_{\kappa}(\mathbf{x},\mathbf{y}),
\end{equation*}
which shows that conformal deformation induces a multiplicative rescaling of the base geodesic distance determined by the values of the conformal factor at the end nodes. This approximation can be interpreted as a discretization of the metric deformation along the geodesic, where the spatial variation of $\xi$ is summarized by an average of the curve’s endpoints. Consequently, the effect of the conformal metric results in an effective modification of the local scale of distances, while preserving the global structure of the geodesics in the surrounding space. In the supplementary material, we propose a simple numerical experiment to show the effect of conformal deformation on the metric when we explicitly modify the metric tensor.

\section{Proposed model}\label{sec3}

We propose an extension of the classical latent space model by incorporating multiplicative nodal effects that modify the effective scale of distances. Instead of assuming a homogeneous global metric, we introduce a node-dependent deformation that acts as a local expansion or contraction factor.

Let $\mathbf{z}^{(i)} \in \mathbb{M}_\kappa^d$ be the latent position and $\xi^{(i)} \in \mathbb{R}_+$ the nodal metric effect for node $i$, where $i=1,\ldots, n$. We define the predictor $s_{i,j}$ as:
\begin{equation*}
s_{i,j}
=
\alpha_0
-
\frac{\xi^{(i)} + \xi^{(j)}}{2}
\, d_\kappa(\mathbf{z}^{(i)},\mathbf{z}^{(j)}). 
\end{equation*}

Conditionally, for an undirected binary network $\mathbf{Y} = (y_{i,j})$, the model is:
\begin{equation*}
y_{i,j} \mid \alpha_0,\xi^{(i)},\xi^{(j)},\mathbf{z}^{(i)},\mathbf{z}^{(j)}
\overset{\text{ind}}{\sim}
\operatorname{Bernoulli}\big(\operatorname{expit}(s_{i,j})\big) \quad 1\leq i < j \leq n. 
\end{equation*}

As an illustrative example of the influence of the nodal metric effect, suppose we have only three nodes 
$i=1,2,3$ located at latent positions such that the geodesic distances between them are $d_{1,2}$, $d_{1,3}$, and $d_{2,3}$. In this case, the model produces three predictors:
\begin{equation*}
s_{1,2} = \alpha_0 - \frac{\xi^{(1)}+\xi^{(2)}}{2} d_{1,2}, \qquad
s_{1,3} = \alpha_0 - \frac{\xi^{(1)}+\xi^{(3)}}{2} d_{1,3}, \qquad
s_{2,3} = \alpha_0 - \frac{\xi^{(2)}+\xi^{(3)}}{2} d_{2,3}.
\end{equation*}

The connection probabilities are then obtained using the function 
$p_{i,j}=\operatorname{expit}(s_{i,j})$. In this formulation, the 
parameters $\xi^{(i)}$ act as multiplicative factors that modify the effective scale of the distances. Large values of $\xi^{(i)}$ ($\xi^{(i)}>1$) expand the distances associated with node $i$, reducing the probability of connection with other nodes, while small values ($0<\xi^{(i)}<1$) contract these distances and favor the formation of links. To visualize this effect, we consider a simple example in which 
we set $\alpha_0=1$ and $d_{1,2}=d_{1,3}=d_{2,3}=1$, and explore different 
configurations of the parameters $(\xi^{(1)},\xi^{(2)},\xi^{(3)})$. Even in this minimal case, different combinations of nodal effects produce qualitatively distinct connection probability patterns, illustrating theadditional flexibility introduced by the model’s local metric deformation.

\begin{figure}[!htb]
\centering
\includegraphics[width=1.0\textwidth]{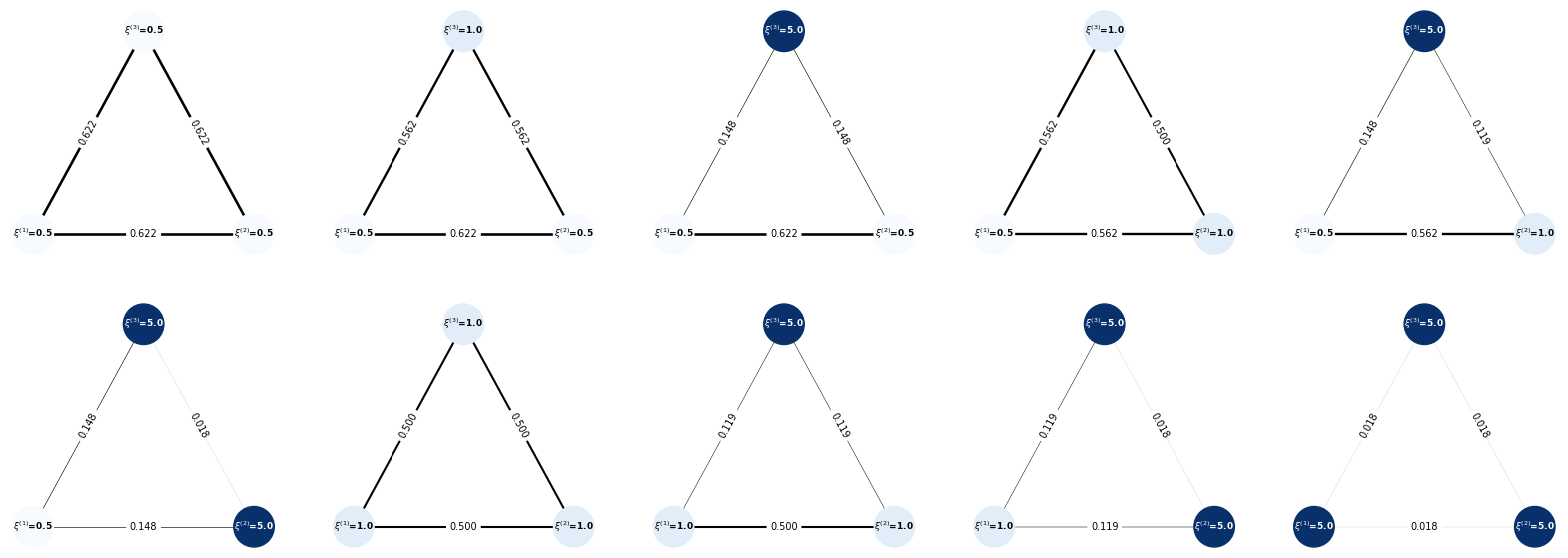}
\caption{Networks generated by the model for different
configurations of the nodal effects. The color and label of the nodes represent the value of $\xi^{(i)}$, while the thickness of the edges and their labels correspond to the connection probability $p_{i,j}$.}
\label{fig:toy_networks}

\end{figure}

The results are summarized in Figure \ref{fig:toy_networks}. Each cell corresponds to a different configuration of the nodal effects $(\xi^{(1)},\xi^{(2)},\xi^{(3)})$, while the links show the values of $p_{1,2}$, $p_{1,3}$, and $p_{2,3}$, which represent the connection probabilities induced by the model. In all cases, $\alpha_0=1$ and $d_{1,2}=d_{1,3}=d_{2,3}=1$ were set.

The figure shows how nodal effects modify the strength of connections even when latent distances remain constant.
When the three nodes have similar values of $\xi^{(i)}$, as in the cases (0.5, 0.5, 0.5), $(1, 1, 1)$, or $(5, 5, 5)$, the connection probabilities are equal among all pairs of nodes, resulting in homogeneous networks. In contrast, when one of the nodes has a significantly higher value of $\xi^{(i)}$, the edges incident to that node have lower probabilities. For example, in the configuration (0.5, 0.5, 5), the probabilities $p_{1,3}$ and $p_{2,3}$ are substantially lower than $p_{1,2}$, reflecting a local increase in the effective approximation distance associated with node $3$. This behavior becomes more pronounced when several nodes have large values of $\xi^{(i)}$. In the extreme configuration $(5,5,5)$, all connection probabilities are small, which corresponds to a global expansion of the latent distances and, therefore, to a highly dispersed network. These results illustrate how the parameters $\xi^{(i)}$ allow for the introduction of structural heterogeneity into the network without altering the latent~positions.

\subsection{Likelihood}

Under conditional independence, the likelihood function is given by $\mathcal{L}_{\kappa}:\mathbb{R}^{1}\times \mathbb{R}^{(d+1)\times n} \times \mathbb{R}^{n}_{+} \to [0,1]\subset\mathbb{R}$, defined as follows:
\begin{equation*}
\mathcal{L}_{\kappa}(\alpha_0,\mathbf{Z},\boldsymbol{\xi}) = p(\mathbf{Y}\mid\alpha_0,\mathbf{Z},\boldsymbol{\xi})
=
\prod_{i,j\,:\,i<j}
p_{i,j}^{y_{i,j}}
(1-p_{i,j})^{1-y_{i,j}},
\end{equation*}
where $p_{i,j}=\operatorname{expit}(s_{i,j})$ and $s_{i,j} = s_{i,j}(\alpha_0,\mathbf{Z},\boldsymbol{\xi})= \alpha_0 - \frac{\xi^{(i)} + \xi^{(j)}}{2} \, d_\kappa(\mathbf{z}^{(i)},\mathbf{z}^{(j)})$. The parameter $\kappa\in \mathbb{R}$ determines the geometry of the latent space, which corresponds to the three simply connected geometric models with constantcurvature: spherical ($\kappa>0$), Euclidean ($\kappa=0$), and hyperbolic ($\kappa<0$). In practice, it is easier to work with the log-likelihood: 
\begin{equation*}
\ell_{\kappa}(\alpha_0,\mathbf{Z},\boldsymbol{\xi}) = \log p(\mathbf{Y}\mid\alpha_0,\mathbf{Z},\boldsymbol{\xi}) = \sum_{i,j\,:\,i<j}
\left[
y_{i,j}s_{i,j}
-
\log(1+e^{s_{i,j}})
\right]. 
\end{equation*}
To perform inference by optimization or using differential information, we present the first-order derivatives of the log-likelihood:

\begin{itemize}
\item Derivative with respect to $\alpha_0$: 
$$
\frac{\partial \ell_{\kappa}}{\partial \alpha_0}=\sum_{i,j\,:\,i<j} (y_{i,j}-p_{i,j}).
$$
\item Derivative with respect to the parameter $\xi^{(m)}$: 
$$
\frac{\partial \ell_{\kappa}}{\partial \xi^{(m)}}=-\frac{1}{2}\sum_{i\,:\,i\neq m}(y_{i,m}-p_{i,m}) d_\kappa(\mathbf{z}^{(i)},\mathbf{z}^{(m)}).
$$
\item Derivative with respect to the latent position $\mathbf{z}^{(m)}$: 
$$
\frac{\partial \ell_{\kappa}}{\partial \mathbf{z}^{(m)}} = -\frac{1}{2} \sum_{i\,:\,i\neq m} (y_{i,m}-p_{i,m}) (\xi^{(i)}+\xi^{(m)})
\frac{\partial d_\kappa(\mathbf{z}^{(i)},\mathbf{z}^{(m)})}{\partial \mathbf{z}^{(m)}}.
$$
\end{itemize}

\subsection{Identifiability}\label{sec:identifiability}

The introduction of multiplicative nodal effects and the possibility of different latent geometries raise potential issues of identifiability. The model depends on the geodesic distances between latent positions and on the nodal parameters $\boldsymbol{\xi}$, which introduces several structural invariants.
Recall that the model’s predictor is $s_{i,j}=\alpha_0-\frac{\xi^{(i)}+\xi^{(j)}}{2}d_\kappa(\mathbf{z}^{(i)},\mathbf{z}^{(j)})$. The log-likelihood depends solely on the predictors $s_{i,j}$, so any transformation of the parameters that preserves these values yields the same likelihood value. Below, we present the model’s invariances and our proposal for addressing them:

\begin{itemize}

\item \textbf{Isometric invariance.}
For any isometry $\operatorname{T}$ of the geometric space $(\mathbb{M}_\kappa^d,g_\kappa)$, we have $d_\kappa(\operatorname{T}\mathbf{z}^{(i)},\operatorname{T}\mathbf{z}^{(j)})=d_\kappa(\mathbf{z}^{(i)},\mathbf{z}^{(j)})$, so the predictors $s_{i,j}$ remain invariant if we hold $\alpha_0$ and $\boldsymbol{\xi}$ fixed; consequently, the latent positions are only identified up to isometric transformations of the latent space.

\item \textbf{Scale indeterminacy.} 
There is an interaction between the scale of the latent positions and the nodal parameters. Indeed, if the distances are scaled as $d_\kappa(\mathbf{z}^{(i)},\mathbf{z}^{(j)}) \mapsto c\, d_\kappa(\mathbf{z}^{(i)},\mathbf{z}^{(j)})$, and simultaneously $\boldsymbol{\xi} \mapsto \frac{1}{c}\boldsymbol{\xi}$ by a constant $c\neq 0$, then the product $\frac{\xi^{(i)}+\xi^{(j)}}{2}d_\kappa(\mathbf{z}^{(i)},\mathbf{z}^{(j)})$ remains invariant, and therefore the predictors $s_{i,j}$ do not change. This implies that the global scale of the latent positions is not uniquely identified. The supplementary material provides a derivation for each of the geometries of non-identifiability due to isometric indeterminacy and also presents the Procrustes problem in order to establish a reference point and solve this problem.

To resolve this indeterminacy, we impose the constraint $\frac{1}{n}\sum_{i=1}^n \xi^{(i)} = 1$. This constraint eliminates the freedom for joint rescaling between $\boldsymbol{\xi}$ and the distances, allowing for a unique identification of the parameters in relative terms; furthermore, this choice also fixes the global scale of the model and allows the parameters $\xi^{(i)}$ to be interpreted as deformation factors relative to a reference geometry. In particular, values of $\xi^{(i)} > 1$ correspond to a local expansion of the metric, while $\xi^{(i)} < 1$ indicate contraction, as evidenced in previous experiments. This constraint is not unique, but it is interpretable from a geometric standpoint, since it preserves the average scale of the latent space and separates the global variation from the local heterogeneity induced by nodal effects. Alternatives such as fixing a norm or a reference value would yield models that are equivalent from a likelihood perspective, but with different interpretations of the parameters.

\end{itemize}

Given these considerations, the model is identifiable modulo the isometries of the latent space and the scale ambiguity described above, both of which are inherent in distance-based models.

An important aspect of the model is that the latent positions $\mathbf{z}^{(i)}$ belong to the geometric manifold $\mathbb{M}_\kappa^d \subset\mathbb{R}^{d+1}$, which depends on the type of geometry considered (spherical, Euclidean, or hyperbolic). As a result, the variables $\mathbf{z}^{(i)}$ are not free parameters in $\mathbb{R}^{d+1}$ but are restricted to a differentiable submanifold. This introduces both theoretical and computational difficulties when performing inference, since the gradients calculated in the ambient space $\mathbb{R}^{d+1}$ do not necessarily belong to the tangent space of the manifold; therefore, a direct update using optimization or sampling methods could produce points outside the latent manifold. Therefore, it is necessary to use optimization methods on Riemannian manifolds \citep{smith2014optimization, absil2008optimization}, where the update directions are projected onto the tangent space and subsequently transported back to the manifold via the exponential map (or approximations thereof); in Section \ref{sec4}, we formalize these ideas derived from optimization on manifolds.

Another important aspect concerns the lack of distance specifications for some of the manifolds under consideration. To resolve this ambiguity, we introduce a geometric constraint that fixes the effective scale of the latent space by imposing a maximum diameter $D > 0$. That is, the geodesic distance between any pair of points must satisfy $d_\kappa(\mathbf{z}^{(i)},\mathbf{z}^{(j)}) \leq D$.
It is important to note that the value of $D$ is not estimated from the data but is set a priori by the modeler. This decision stems from the fact that one of the central objectives of the approach is to compare the model’s performance under different geometries (spherical, Euclidean, and hyperbolic) on a common scale. A joint inference of this diameter would require learning the three considered geometries in a coupled manner, thereby obscuring the objective of evaluating their differential effects under controlled conditions.
Note that this condition is naturally compatible with the spherical case, where the diameter is intrinsically bounded by the geometry, whereas in the Euclidean and hyperbolic cases it constitutes an additional restriction that can be interpreted as an effective compactification of the latent space. Below we discuss the restriction for each of the ambient spaces considered in this work:

\begin{itemize}

\item \textbf{Euclidean case ($\kappa = 0$).}
In $\mathbb{M}_0^d \cong \mathbb{R}^d$, the geodesic distance coincides with the Euclidean norm. The diameter restriction $D$ implies that the latent positions must belong to a closed ball of radius $D$, that is, $\|\mathbf{z}^{(i)} - \mathbf{z}^{(j)}\| \leq D$ for all $i,j$. An equivalent way to impose this restriction is to require that $\mathbf{z}^{(i)} \in B(\mathbf{0}, D/2)$ for all $i$, where $B(\mathbf{0}, D/2)$ denotes the Euclidean ball centered at the origin with radius $D/2$.

\item \textbf{Spherical case ($\kappa > 0$).}
In $\mathbb{M}_\kappa^d = \mathbb{S}^d$, the geodesic distance is bounded above by $\pi R$, where $R = \frac{1}{\sqrt{\kappa}}$ is the radius of the sphere. Therefore, the maximum diameter of the space is $\pi R$. In this sense, if we want to impose a maximum distance of $D$ between two points, it suffices to set $R = \frac{D}{\pi}$, or, in terms of the curvature, $\kappa = \frac{\pi^2}{D^2}$.

\item \textbf{Hyperbolic case ($\kappa < 0$).}
In $\mathbb{M}_{\kappa}^d = \mathbb{H}^d$, the space is noncompact and the geodesic distance is not bounded above. Therefore, the diameter constraint $D$ is implemented by restricting the latent positions to a geodesic ball of radius $D/2$ centered at a reference point. In particular, we set the center to the point $R\,\mathbf{e}_{d+1}$ on the hyperboloid model, where $\mathbf{e}_{d+1} = (0,\dots,0,1)^\top \in\mathbb{R}^{d+1}$ and $R = \frac{1}{\sqrt{-\kappa}} > 0$ satisfies $\langle R\,\mathbf{e}_{d+1}, R\,\mathbf{e}_{d+1} \rangle_L = -R^2$. The constraint is expressed as $d_{\kappa}(\mathbf{z}^{(i)}, R\,\mathbf{e}_{d+1}) \leq \frac{D}{2}$ or, equivalently, $z_{d+1}^{(i)}\leq R\operatorname{cosh}\left (\frac{D}{2R}\right)$ for all $i$, which guarantees that $d_{\kappa}(\mathbf{z}^{(i)}, \mathbf{z}^{(j)}) \leq D$ for all $i,j$. Note that in the hyperbolic case, there are two free parameters that determine the geometry: the intrinsic radius $R$ (equivalently, the curvature $\kappa$) and the effective diameter $D$, which restricts the region of latent space. This redundancy reflects an additional scale indeterminacy in the model.
In order to obtain an identifiable parameterization and facilitate comparisons between geometries with positive, zero, and negative curvature, we establish a relationship between these parameters by setting $R := \frac{D}{\pi}$. Under this choice, the magnitude of thecurvature is normalized as $|\kappa| = \pi^2 / D^2$, which allows us to consider a symmetric family of models withcurvatures $\kappa \in \{-C, 0, C\}$ for a constant $C > 0$. This choice is not unique, but it is convenient because it decouples the variation in curvature from the global scale effect and provides a homogeneous parameterization across different geometric~regimes.
\end{itemize}

\subsection{Prior and posterior distribution}

\begin{figure}[!htb]
\centering
\includegraphics[width=0.9\linewidth]{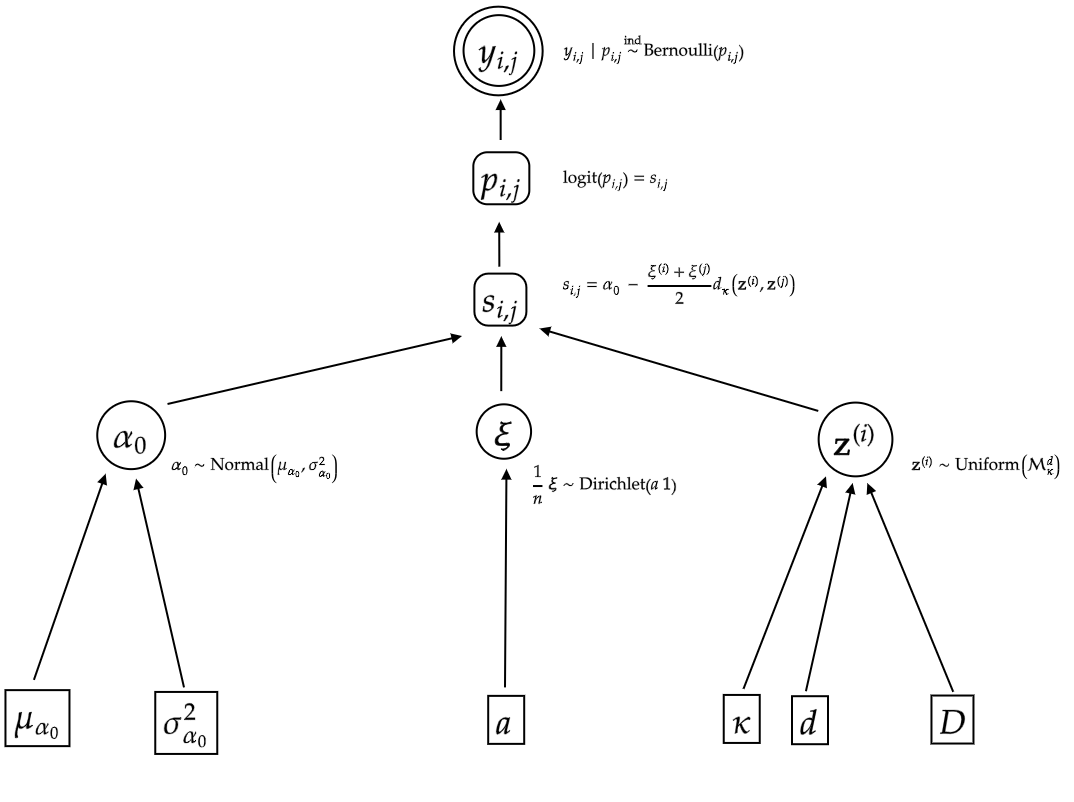}
\caption{Directed acyclic graph (\textit{DAG}) for the proposed model. Variables with a double circle represent observed data; variables with a circle are considered random variables; variables with a square with rounded corners are deterministic terms obtained for a particular realization of the random variables; and variables with rectangles are fixed model parameters. The directed arrows represent dependencies between variables, with the arrowhead pointing to the variable that depends on the variable at the tail of the arrow.}
\label{fig:DAG}
\end{figure}

Let $\mathbf{Y} = (y_{i,j}) \in \mathbb{R}^{n \times n}$ be the adjacency matrix of an observed undirected binary network. In accordance with the hierarchical structure of the model (represented by the DAG in Figure \ref{fig:DAG}), the probabilities of interaction between pairs of nodes are constructed from geometric latent variables and nodal effects. The probability of a connection between nodes $i$ and $j$ is given by:
\begin{equation*}
p_{i,j} = \operatorname{expit}(s_{i,j}), 
\quad \text{where} \quad 
s_{i,j} = \alpha_0 - \frac{\xi^{(i)} + \xi^{(j)}}{2}\, d_\kappa(\mathbf{z}^{(i)}, \mathbf{z}^{(j)}),
\end{equation*}
where $\kappa \in \{-C, 0, C\}$ and $C = \frac{\pi^2}{D^2}$ determine the underlying geometry of the latent space. This specification directly reflects the geometric interpretation of the model: the link probability decreases with geodesic distance, modulated by nodal effects that act as local scaling factors. Given the latent parameters $(\alpha_0, \mathbf{Z}, \boldsymbol{\xi})$, the edges are assumed to be independent. This assumption, which is standard in latent space models, is consistent with the factorization induced by the DAG and allows us to write:
\begin{equation*}
y_{i,j} \mid \alpha_0, \mathbf{Z}, \boldsymbol{\xi} 
\;\overset{\text{ind}}{\sim}\; \operatorname{Bernoulli}(p_{i,j}), 
\quad 1 \leq i < j \leq n.
\end{equation*}
Thus, all the interdependence observed in the network is introduced indirectly through the latent structure.

\paragraph{Prior distributions.}

The Bayesian specification of the model is completed by assigning prior distributions to the latent parameters, while preserving the hierarchical structure. First, the latent positions $\mathbf{z}^{(i)}$, for $i=1,\ldots,n$, are assumed to be independent and uniformly distributed over the manifold $\mathbb{M}_\kappa^d$ restricted to a maximum diameter $D$. 

This choice reflects the absence of prior information about the location of the nodes in latent space, while maintaining the model’s geometric invariance. Meanwhile, the nodal effects $\boldsymbol{\xi} = (\xi^{(1)},\ldots,\xi^{(n)})^{\top}$ are modeled using a Dirichlet distribution $\frac{1}{n}\boldsymbol{\xi} \sim \operatorname{Dirichlet}(a \mathbf{1}_n)$, where $a$ is the concentration parameter. The global intercept $\alpha_0$ controls the average density of the network and is modeled using a normal distribution $\alpha_0 \sim \operatorname{Normal}(\mu_{\alpha_0}, \sigma_{\alpha_0}^2)$, where $\mu_{\alpha_0}$ is a fixed parameter representing the mean of the distribution and $\sigma_{\alpha_0}^2$ is a fixed variance parameter.

Below, we explicitly state the log-densities for each of the model’s parameters. The latent positions have a density that can be written as $p(\mathbf{Z}) = \prod_{i=1}^{n}p(\mathbf{z}^{(i)})$, where each $p(\mathbf{z}^{(i)})$ is equal to the multiplicative inverse of the volume of the latent space; therefore, the quantity $\log(p(\mathbf{Z}))$ is constant. Meanwhile, the node effects $\boldsymbol{\xi} = (\xi^{(1)},\ldots,\xi^{(n)})^{\top}$ are modeled using a Dirichlet distribution
$\frac{1}{n}\boldsymbol{\xi} \sim \operatorname{Dirichlet}(a \mathbf{1}_n)$, whose density is given by $p\left (\frac{1}{n}\boldsymbol{\xi}\right) = \frac{\Gamma(na)}{\Gamma(a)^n} \prod_{i=1}^n \left(\frac{1}{n}\xi^{(i)}\right)^{a-1}$ and therefore its logarithmic form is:
\begin{equation*}
\log p\left(\frac{1}{n}\boldsymbol{\xi}\right) 
= \log \Gamma(na) - n \log \Gamma(a) + (a-1)\sum_{i=1}^n \log\left(\frac{1}{n} \xi^{(i)}\right).
\end{equation*}

The global intercept $\alpha_0$ is modeled as $\alpha_0 \sim \operatorname{Normal}(\mu_{\alpha_0}, \sigma_{\alpha_0}^2)$, whose log-density is:
\begin{equation*}
\log p(\alpha_0)
= -\frac{1}{2}\log(2\pi \sigma_{\alpha_0}^2)
-\frac{1}{2\sigma_{\alpha_0}^2}(\alpha_0 - \mu_{\alpha_0})^2.
\end{equation*}

\paragraph{Posterior distribution}

Given the full hierarchical model, the joint posterior distribution must include all the parameters and latent variables of the model, that is, $\Theta = (\alpha_0, \mathbf{Z}, \boldsymbol{\xi})$. Therefore, the posterior is obtained using Bayes' rule as follows:
\begin{equation*}
p(\Theta \mid \mathbf{Y}) \propto p(\mathbf{Y} \mid \Theta) \;
p(\Theta). 
\end{equation*}
The likelihood is factored into a product of Bernoulli terms due to conditional independence:
\begin{equation*}
 p(\mathbf{Y} \mid \Theta) = p(\mathbf{Y} \mid \alpha_0, \mathbf{Z}, \boldsymbol{\xi})
= \prod_{i,j\,:\,i<j} p_{i,j}^{y_{i,j}} (1-p_{i,j})^{1-y_{i,j}}, 
\end{equation*}
while the hierarchical structure leads to the following decomposition of the density of the prior distribution:
\begin{equation*}
p(\Theta) = p(\mathbf{Z}) \, p\left(\frac{1}{n}\boldsymbol{\xi} \right)\, p(\alpha_0).\end{equation*}
In logarithmic terms, the density of the posterior distribution (up to an additive constant) can be written as:
\begin{align*}
\log p(\Theta \mid \mathbf{Y}) 
&\propto \sum_{i,j\,:\,i<j} \left[
y_{i,j} s_{i,j} - \log\big(1 + e^{s_{i,j}}\big)
\right] + \log p(\mathbf{Z})
+ \log p\left(\frac{1}{n}\boldsymbol{\xi}\right)
+ \log p(\alpha_0).
\end{align*}

Additive decomposition on a logarithmic scale is useful for designing inference algorithms, since it allows blocks of parameters to be updated separately.

\subsection{Generative model}

To evaluate the flexibility and expressive power of the proposed latent space model with metric deformation, we study its generative behavior in controlled simulations.
Following the experimental strategy outlined in \cite{smith2019}, we analyze how the
underlying geometry and the inclusion of node-specific metric effects influence the structural properties of the generated networks. In the first place, we compare the impact of nodal effects on the distance between a pair of latent positions and also how this is reflected in connection probabilities and the degree distribution. 

\begin{figure}[!htb]
\centering
\hspace*{0.75cm}\includegraphics[width=0.905\linewidth]{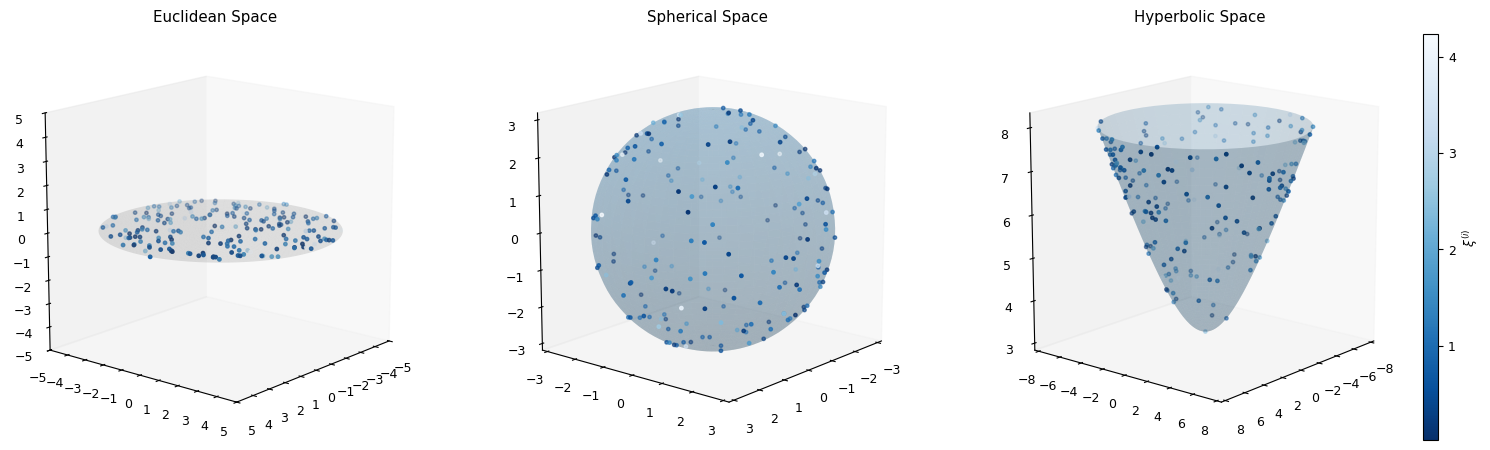}
\includegraphics[width=0.905\linewidth]{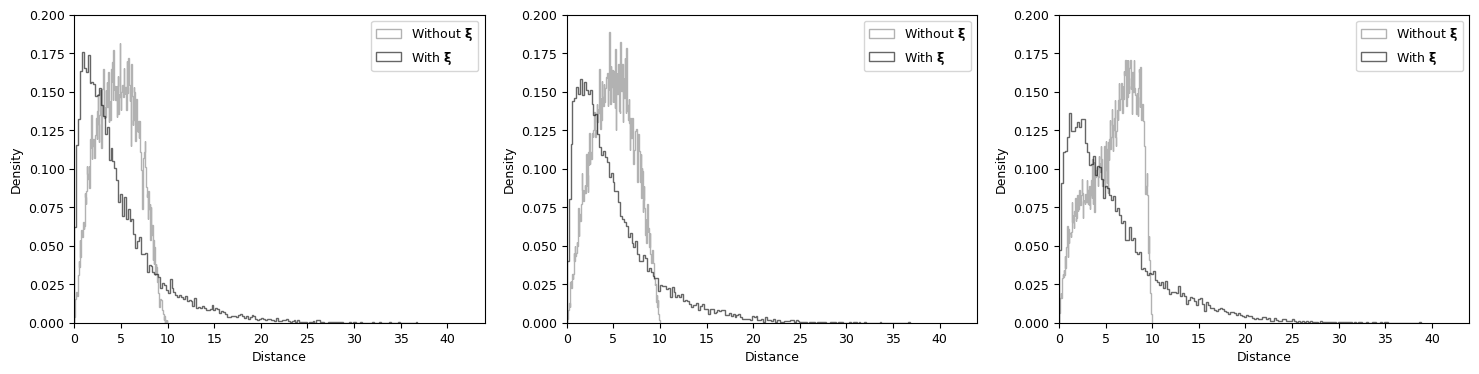}
\caption{The top row shows the three latent space geometries ($\mathbb{R}^2$, $\mathbb{S}^2$, and $\mathbb{H}^2$, restricted to a diameter of $D=10$) embedded in $\mathbb{R}^3$, along with 200 uniformly sampled points colored according to the value of $\xi^{(i)}$. The bottom row shows the histogram of the distances between pairs of latent positions, separated by geometry and by the presence or absence of the multiplicative effects~$\boldsymbol{\xi}$.}
\label{fig:exp4_geometries}
\end{figure}

\begin{figure}[!htb]
\centering
\begin{minipage}{0.47\textwidth}
\centering
\includegraphics[width=\linewidth]{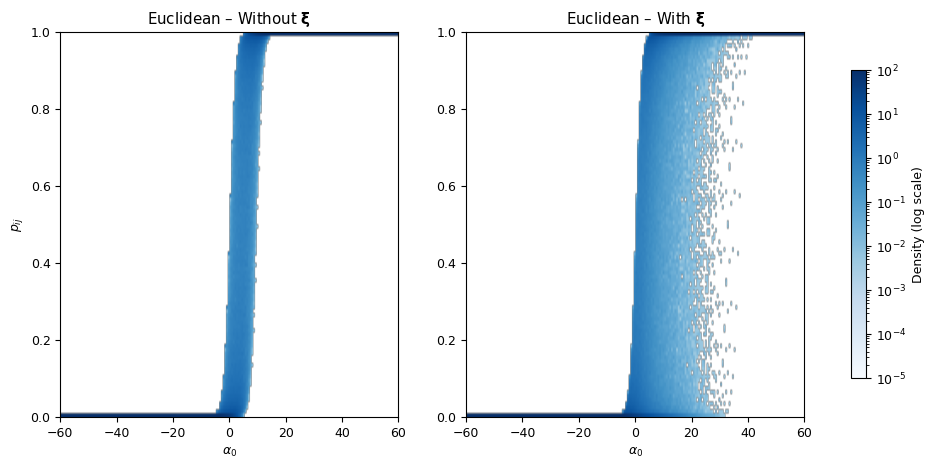}
\captionof*{figure}{(a) Distribution of the value of $p_{i,j}$ as $\alpha_0$ varies for Euclidean geometry.}
\end{minipage} 
\hfill
\begin{minipage}{0.51\textwidth}
\centering
\includegraphics[width=\linewidth]{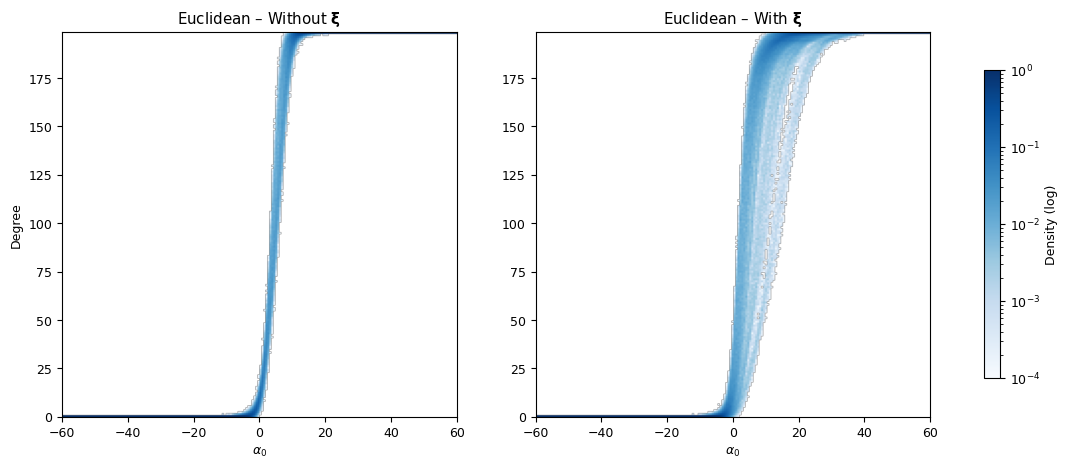}
\captionof*{figure}{(b) Distribution of the degree as $\alpha_0$ varies for Euclidean geometry.}
\end{minipage}
\caption{Visualization of the impact of multiplicative effects on distance for the connection probability distribution and for the degree as $\alpha_0$ varies. For each value of $\alpha_0$, a histogram is calculated and represented in the figures by color.}
\label{fig:exp4_alpha0}
\end{figure}

For each geometry (hyperbolic, Euclidean, and spherical), we consider a single realization of the latent positions $\{\mathbf{z}^{(i)}\}_{i=1}^n$ from the base manifold $\mathbb{M}_\kappa^d$ with a maximum diameter constraint of $D$. Conditionally on this sample, networks are generated according to the classical latent space model ($\boldsymbol{\xi} = \mathbf{1}_n$) and the proposed model with metric deformations at the node level, where $\frac{1}{n}\boldsymbol{\xi}\sim \mathrm{Dirichlet}(\mathbf{1}_n)$. Figure \ref{fig:exp4_geometries} shows the sampled latent positions alongside the resulting distance distribution. Across the three geometries, a common behavior is observed when multiplicative effects are introduced into the distances: when multiplicative effects are present, the distance distribution has a larger support and exhibits positive asymmetry, indicating a greater concentration of small distances and a few large distances. Furthermore, Figure \ref{fig:exp4_alpha0} allows us to analyze how the metric deformation induced by the nodal effects $\boldsymbol{\xi}$ interacts with the global parameter $\alpha_0$ in the formation of links with respect to the probability $p_{i,j}$.

In the absence of multiplicative effects ($\boldsymbol{\xi} = \mathbf{1}_n$), the distribution of $p_{i,j}$ for all pairs of nodes exhibits a relatively concentrated transition as $\alpha_0$ varies. For small values of $\alpha_0$, the connection probabilities are concentrated near zero, whereas for large values they are concentrated near one, with a narrow transition zone. This reflects the fact that, under a homogeneous metric, the variability in $p_{i,j}$ is determined exclusively by the distribution of latent distances, which remains fixed. In contrast, when multiplicative effects ($\frac{1}{n}\boldsymbol{\xi} \sim \mathrm{Dirichlet}(\mathbf{1}_n)$) are introduced, the transition in the distribution of $p_{i,j}$ becomes considerably more diffuse. For a given value of $\alpha_0$, greater dispersion is observed in the values of $p_{i,j}$, simultaneously covering regions close to both 0 and 1. This phenomenon is a direct consequence of the heterogeneity induced in the effective distances: pairs of nodes with small values of $\xi^{(i)} + \xi^{(j)}$ exhibit contracted distances and, therefore, high connection probabilities, while pairs with large values experience an expansion of the distance and low probabilities. In geometric terms, the conformal deformation breaks the local homogeneity of the latent space, generating a node-dependent “rescaling”.

This behavior has a direct impact on the degree distribution, as shown in Figure \ref{fig:exp4_alpha0}(b). In the classical model, the degree distribution evolves in a relatively concentrated manner as $\alpha_0$ increases, transitioning uniformly from sparse networks to dense networks. In contrast, under the model with nodal effects, the degree distribution exhibits significantly greater dispersion for intermediate values of $\alpha_0$. In particular, nodes with very low degrees coexist with highly connected nodes within the same network. This broadening of the degree distribution is a direct manifestation of the structural heterogeneity induced by $\boldsymbol{\xi}$. Nodes with small values of $\xi^{(i)}$ act as “hubs” or attractors by shortening their distances to the rest of the nodes, while nodes with large values of $\xi^{(i)}$ become peripheral by increasing their distances. Unlike the classical model, where degree heterogeneity is explained primarily by global geometry, here it emerges locally even in homogeneous geometries such as Euclidean~geometry.

As additional evidence, the supplementary material includes three additional experiments on the effect of the Dirichlet prior distribution on nodal effects, as well as a comparison of different network statistics for networks simulated using the classical model and the proposed model, and the change in the Laplacian spectrum when multiplicative nodal effects are incorporated.

\section{Inference and computation}\label{sec4}

Estimating the parameters of the proposed model requires simultaneously addressing continuous latent variables, nodal effects, and the underlying geometry. The hierarchical structure of the model is summarized in the directed acyclic graph (DAG) shown in Figure~\ref{fig:DAG}, which explicitly depicts the dependencies between observed variables and latent parameters. The computational complexity stems primarily from the repeated calculation of geodesic distances and the high dimensionality of the latent positions.

\subsection{Inference by optimization}

The optimization-based estimation in the proposed model requires optimizing the log-posterior
over a mixed parameter space: the global intercept $\alpha_0$ belongs to Euclidean space,
the nodal effects $\boldsymbol{\xi}$ reside in the scaled simplex $\mathcal C(n)=\left\{\boldsymbol{x}\in\mathbb R_+^n:\sum_{i=1}^n x_i=n\right\}$,and the latent positions $\mathbf Z=[\mathbf z^{(1)},\cdots,\mathbf z^{(n)}]$ belong to the product of
$(\mathbb M_\kappa^d)^n \subset \mathbb{R}^{(d+1)\times n}$ with maximum diameter $D$, where $\mathbb M_\kappa^d$ can be Euclidean,
spherical, or hyperbolic depending on the value of $\kappa$. Consequently, optimization cannot be performed using ordinary Euclidean gradients, but must respect the intrinsic geometry of each space.

Let $\mathbf{z}\in\mathbb{M}_{\kappa}^{d}$ be a point on a Riemannian manifold, and let
$T_\mathbf{z}\mathbb{M}_{\kappa}^{d}$ be its tangent space. Given a tangent vector $\mathbf{v}\in T_{\mathbf{z}}\mathbb{M}_{\kappa}^{d}$,
the exponential map at $\mathbf{z}$ is defined as $\operatorname{Exp}_{\mathbf{z}}(\mathbf{v}) =\gamma_{\mathbf{v}}(1)$, where $\gamma_{\mathbf{v}}$ is the geodesic satisfying $\gamma_{\mathbf{v}}(0)=\mathbf{z}$ and $\dot\gamma_{\mathbf{v}}(0)=\mathbf{v}$. This operator transports a tangent direction back to the manifold and constitutes the Riemannian generalization of the additive step $\mathbf{z}+\mathbf{v}$ in Euclidean spaces \citep{smith2014optimization, absil2008optimization}.

\begin{itemize}
\item In Euclidean space $\mathbb{R}^d$, geodesics are straight lines, so the exponential map coincides with the ordinary vector addition:
\begin{equation*}
\operatorname{Exp}_{\mathbf z}(\mathbf v)=\mathbf z+\mathbf v.
\end{equation*}
\item For the sphere $\mathbb S^d$ of radius $R$, the exponential map on $\mathbf z\in\mathbb S^d$ is given by:
\begin{equation*}
\operatorname{Exp}_{\mathbf z}(\mathbf v)
=
\cos\!\left(\frac{\|\mathbf v\|}{R}\right)\mathbf z
+
R\sin\!\left(\frac{\|\mathbf v\|}{R}\right)
\frac{\mathbf v}{\|\mathbf v\|},
\qquad \mathbf v\in T_{\mathbf z}\mathbb S^d.
\end{equation*}
This expression moves the point along the great circle determined by the tangent direction $\mathbf v$.
\item In the model of the hyperboloid $\mathbb H^d$ of intrinsic radius $R$, the exponential map takes the form:
\begin{equation*}
\operatorname{Exp}_{\mathbf z}(\mathbf v)
=
\cosh\!\left(\frac{\|\mathbf v\|_L}{R}\right)\mathbf z
+
R\sinh\!\left(\frac{\|\mathbf v\|_L}{R}\right)
\frac{\mathbf v}{\|\mathbf v\|_L},
\qquad \mathbf v\in T_{\mathbf z}\mathbb H^d,
\end{equation*}
where $\|\mathbf v\|_L=\sqrt{\langle \mathbf v,\mathbf v\rangle_L}$ is the norm induced by the Lorentz product.
\end{itemize}

Before applying the exponential map, the gradients must be projected onto the tangent space.
If $\mathbf{v} = \nabla f(\mathbf{z})$ denotes the Euclidean gradient in the ambient space for a scalar function $f$ defined on the manifold, its projection is $\operatorname{Proj}_{T_{\mathbf{z}}\mathcal M}(\nabla f(\mathbf{z}))$:

\begin{itemize}
\item In Euclidean geometry, the tangent projection is trivial $\operatorname{Proj}_{T_{\mathbf{z}}\mathbb{M}_{\kappa}^{d}}(\mathbf{v})=\mathbf{v}$.

\item In spherical geometry, the projection is $\operatorname{Proj}_{T_{\mathbf{z}}\mathbb{M}_{\kappa}^{d}}(\mathbf{v})
=\mathbf{v}-\frac{\langle {\mathbf{z}},\mathbf{v}\rangle}{R^2}\mathbf{z}$.
\item In hyperbolic geometry, represented by the hyperboloid model, the projection corresponds to $\operatorname{Proj}_{T_{\mathbf{z}}\mathbb{M}_{\kappa}^{d}}(\mathbf{v})
=\mathbf{v}+\frac{\langle \mathbf{z},\mathbf{v}\rangle_L}{R^2}\mathbf{z}$,
where $\langle\cdot,\cdot\rangle_L$ is the Lorentz product.
\end{itemize}

After each update, a projection is also applied to the admissible domain to ensure that the parameters remain within the restricted range or set:

\begin{itemize}
\item \textbf{Projection onto the simplex.}
For the nodal vectors$\boldsymbol{\xi}\in\mathbb{R}^n$, the projection onto the simplex $\mathcal C(n)$is performed in two steps: first, positivity is enforced component by component, $\tilde\xi^{(i)}=\max(\xi^{(i)},\varepsilon)$ for $i=1,\dots,n$, where $\varepsilon>0$ is a small numerical stability constant; it is then renormalized $\operatorname{Proj}_{\mathcal C(n)}(\boldsymbol{\xi};\varepsilon)
=
\frac{n}{\sum_{j=1}^n\tilde\xi^{(j)}}
\,\tilde{\boldsymbol{\xi}}.$

\item \textbf{Projection in the bounded Euclidean case.}
If $\mathbb M_\kappa^d$ represents the Euclidean space with diameter restriction $D$, the admissible domain is the closed Euclidean ball of radius $D/2$ centered at the origin. Given $\mathbf z\in\mathbb R^d$, we define:
\begin{equation*}
\operatorname{Proj}_{\mathbb M_\kappa^d }(\mathbf z)=
\begin{cases}
\mathbf z, & \|\mathbf z\|\le D/2,\\[1ex]
\dfrac{D/2}{\|\mathbf z\|}\mathbf z, & \|\mathbf z\|>D/2.
\end{cases}
\end{equation*}
In other words, if the point falls outside the sphere, it is rescaled radially to the boundary while maintaining its direction.

\item \textbf{Projection onto the sphere.}
In the spherical manifold $\mathbb M_\kappa^d$, any point in the ambient space $\mathbb R^{d+1}$ is projected radially onto the sphere:
\begin{equation*}
\operatorname{Proj}_{\mathbb M_\kappa^d}(\mathbf z)
=
R\frac{\mathbf z}{\|\mathbf z\|}. 
\end{equation*}

\item \textbf{Projection onto the hyperbolic hyperboloid.}
In the hyperbolic model $\mathbb M_\kappa^d$, we write
$\mathbf{z}=(z_1,\dots,z_d,z_{d+1})^{\top}=(\mathbf{z}_{1:d}^{\top},z_{d+1})^{\top}$ and restrict the domain to the geodesic ball of radius $D/2$ centered at
$(0,\dots,0,R)$. We define $t_{\max}=R\cosh\!\left(\frac{D}{2R}\right)$. Then the projection is given by:
\begin{equation*}
\operatorname{Proj}_{\mathbb M_\kappa^d}(\mathbf z)=
\begin{cases}
\mathbf{z} & z_{d+1}\le t_{\max},\\[1ex]
\left(
\dfrac{\sqrt{t_{\max}^2-R^2}}{\|\mathbf z_{1:d}\|}\mathbf{z}_{1:d}^{\top},\;
t_{\max}
\right)^{\top}, & {z_{d+1}}>t_{\max}.
\end{cases}
\end{equation*}
\end{itemize}

With these operations, each iteration of the algorithm performs Riemannian gradient ascent \citep{smith2014optimization, absil2008optimization}; that is, the gradient is first computed in the ambient space, then projected onto the tangent space, subsequently transported via the exponential map, and finally reprojected onto the allowed geometric domain. This ensures that all updates respect the geometric structure of the latent model. 

However, in the proposed model, there is a strong dependence between the latent positions ($\mathbf Z$) and the nodal effects ($\boldsymbol{\xi}$). In fact, both parameters simultaneously influence the predictor $s_{i,j}$, such that a single connection probability can be explained by different combinations of geodesic distances and nodal effects. Consequently, the log-posterior surface exhibits relatively flat regions and a high correlation between the two blocks of parameters, hindering joint optimization and increasing sensitivity to initialization.
In order to decouple these two explanatory mechanisms, we adopt a sequential optimization strategy inspired by\emph{block coordinate optimization} methods \citep{meng1993maximum,lange2016mm,razaviyayn2013unified}. The idea is to use the classical latent space model to first obtain a stable estimate of the underlying geometry and then estimate the nodal effects by conditioning on that geometry. For the classical model, we define the estimators as:
\begin{equation*}
({\hat{\mathbf Z},\hat{\alpha}_0}) = \arg\max_{\mathbf Z,\alpha_0}
\left\{\log p(\mathbf Z,\alpha_0\mid\mathbf Y)
\right\},
\end{equation*}
furthermore, for the proposed model, we define the estimator as follows:
\begin{equation*}
\hat{\boldsymbol{\xi}}=\arg\max_{\boldsymbol{\xi}}
\left\{
\log p(\hat{\mathbf Z},\boldsymbol{\xi},\hat{\alpha}_0\mid\mathbf Y)
\right\},
\end{equation*}
that is, the estimators $({\hat{\mathbf Z},\hat{\alpha}_0})$ will be the same for both models, and $\boldsymbol{\xi}$ is estimated by maximizing the log-posterior while fixing the latent positions and the intercept estimated from the classical model. From a statistical perspective, this procedure can be interpreted as a \emph{plug-in} approximation, where the uncertainty associated with the latent positions is replaced by their MAP estimator obtained under the classical model. This strategy reduces the competition between the geometric parameters and the nodal effects, allowing the latter to explain only the residual heterogeneity not captured by the geometry of the latent space. Furthermore, it reduces the effective dimensionality of the optimization problem and leads to a numerically more stable procedure without altering the geometric interpretation of the model.

Algorithms~\ref{alg:classical} and~\ref{alg:plugin} summarize the estimation procedure. Algorithm~\ref{alg:classical} obtains the MAP estimators of the latent positions and intercept through Riemannian gradient ascent with a multi-start strategy, retaining the solution that maximizes the log-posterior. Subsequently, Algorithm~\ref{alg:plugin} estimates the nodal effects by maximizing the conditional log-posterior while keeping the estimated latent positions and intercept fixed. This sequential plug-in strategy separates the estimation of the latent geometry from the estimation of node-specific effects, improving numerical stability and reducing the competition between both sets of parameters.

\begin{algorithm}[H]
\caption{Estimation for the classical latent space model}
\label{alg:classical}
\begin{algorithmic}[1]

\Require Adjacency matrix $\mathbf{Y}$, number of random initializations $K$, maximum number of iterations $T$, learning rate $\eta$.
\Ensure Estimators $(\hat{\mathbf Z},\hat{\alpha}_0)$.

\State $\ell^\star\leftarrow -\infty$

\For{$k=1,\ldots,K$}

\State Sample initial values $\mathbf Z^{[0]}$ and $\alpha_0^{[0]}$ from their prior distributions.

\For{$t=1,\ldots,T$}

\State Compute $g_{\alpha_0}
=
\nabla_{\alpha_0}
\log p(\mathbf Z^{[t-1]},\alpha_0^{[t-1]}\mid\mathbf Y).$
\State Update $\alpha_0^{[t]}
=
\alpha_0^{[t-1]}
+
\eta \, g_{\alpha_0}.$
\State Compute the ambient gradient $ \mathbf G_Z
=
\nabla_{\mathbf Z}
\log p(\mathbf Z^{[t-1]},\alpha_0^{[t]}\mid\mathbf Y).
$

\State Project onto the tangent space
$
\mathbf G_Z
\leftarrow
\operatorname{Proj}_{T_{\mathbf Z}\mathbb M_\kappa^d}
(\mathbf G_Z).
$

\State Update
$
\mathbf Z^{[t]}
=
\operatorname{Proj}_{\mathbb M_\kappa^d}
\left(
\operatorname{Exp}_{\mathbf Z^{[t-1]}}
(\eta\,\mathbf G_Z)
\right).
$

\State Evaluate
$\ell^{[t]}
=
\log p(\mathbf Z^{[t]},\alpha_0^{[t]}\mid\mathbf Y)$.

\If{$\ell^{[t]}>\ell^\star$}
\State Save
$\hat{\mathbf Z}\leftarrow\mathbf Z^{[t]}$,
$\hat{\alpha}_0\leftarrow\alpha_0^{[t]}$,
$\ell^\star\leftarrow\ell^{[t]}$.
\EndIf
\If{early stopping criterion is satisfied}
\State \textbf{break}
\EndIf

\EndFor

\EndFor

\State \Return $(\hat{\mathbf Z},\hat{\alpha}_0)$

\end{algorithmic}
\end{algorithm}

\begin{algorithm}[H]
\caption{Plug-in estimation of nodal effects}
\label{alg:plugin}
\begin{algorithmic}[1]

\Require $\mathbf Y$, estimators $(\hat{\mathbf Z},\hat{\alpha}_0)$ from Algorithm~\ref{alg:classical}, number of random initializations $K$, maximum number of iterations $T$, learning rate $\eta$, positivity constant $\varepsilon$.
\Ensure Estimator $\hat{\boldsymbol{\xi}}$.
\State $\ell^\star\leftarrow -\infty$
\For{$k=1,\ldots,K$}

\State Sample $\boldsymbol{\xi}^{[0]}$ from its prior distribution.

\For{$t=1,\ldots,T$}

\State Compute $\mathbf g_{\xi}
=
\nabla_{\boldsymbol\xi}
\log
p(
\hat{\mathbf Z},
\boldsymbol\xi^{[t-1]},
\hat{\alpha}_0
\mid
\mathbf Y).$

\State Project onto the tangent space of the simplex
$
\mathbf g_{\xi}
\leftarrow
\operatorname{Proj}_{T_{\boldsymbol\xi}\mathcal C(n)}
(\mathbf g_{\xi}).
$

\State Update
$
\boldsymbol\xi^{[t]}
=
\operatorname{Proj}_{\mathcal C(n)}
\left(
\boldsymbol\xi^{[t-1]} +\eta\,\mathbf g_{\xi},
\varepsilon
\right).
$

\State Evaluate
$
\ell^{[t]}
=
\log
p(
\hat{\mathbf Z},
\boldsymbol\xi^{[t]},
\hat{\alpha}_0
\mid
\mathbf Y).
$
\If{$\ell^{[t]}>\ell^\star$}
\State Save $\hat{\boldsymbol\xi}\leftarrow\boldsymbol\xi^{[t]}$,
$\ell^\star\leftarrow\ell^{[t]}$.
\EndIf
\If{early stopping criterion is satisfied}
\State \textbf{break}
\EndIf

\EndFor
\EndFor

\State \Return $\hat{\boldsymbol\xi}$

\end{algorithmic}
\end{algorithm}

\section{Inference about some networks}\label{sec5}

To evaluate the empirical performance of the proposed model, we analyze both synthetic networks and real-world data from various domains, covering different structural regimes. In each case, we systematically compare the classical latent space model with the extension that incorporates multiplicative node effects. The evaluation is conducted in terms of statistical fit, predictive power, and structural consistency. This analysis allows us to quantify the impact of metric deformation on the representation of properties such as centrality, clustering, homophily, and degree heterogeneity. In the supplementary material, we present the results obtained by applying the proposed model and comparing it to the classical model and six other example networks. All code
required to reproduce the analyses is available at the following repository \href{https://github.com/cnosa/ConformalDeformationLSMN}{\texttt{github.com/cnosa/ConformalDeformationLSMN}}.

\subsection{Florentine families}

The network of marriage alliances among the Florentine families of the Renaissance is one of the most extensively studied datasets in social network analysis. Originally introduced by \cite{padgett_robust_1993}, it describes the matrimonial ties established among $n=$15 influential families in fifteenth-century Florence. These alliances played a central role in shaping the political and economic organization of the city, making the network a canonical benchmark for evaluating statistical models of social networks.

\begin{figure}[!htb]
\centering
\begin{minipage}{0.32\textwidth}
\centering
\includegraphics[width=\linewidth]{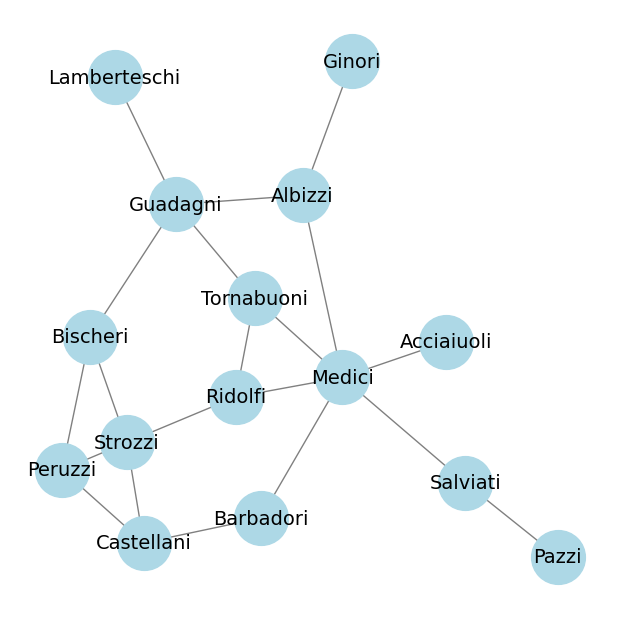}
\captionof*{figure}{(a) Data with~family~names. }
\end{minipage}
\begin{minipage}{0.32\textwidth}
\centering
\includegraphics[width=\linewidth]{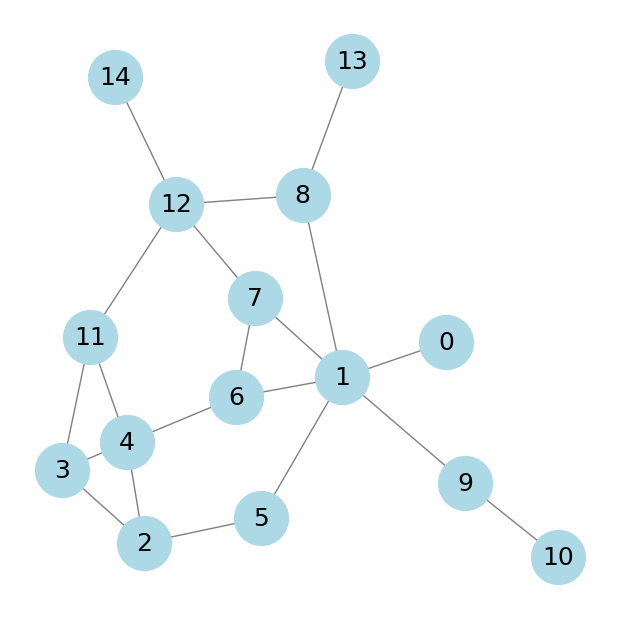}
\captionof*{figure}{(b) Data with labeled nodes. }
\end{minipage}
\begin{minipage}{0.32\textwidth}
\centering
\includegraphics[width=\linewidth]{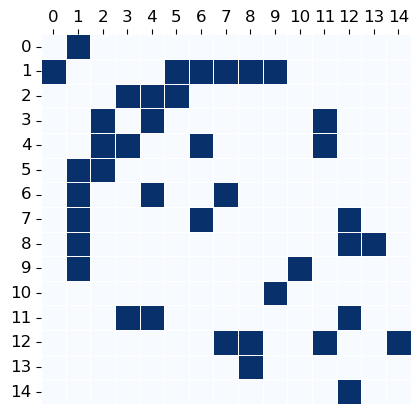}
\captionof*{figure}{(c) Adjacency matrix. }
\end{minipage}
\caption{Visualization of the network of Florentine families.}
\label{fig:Florentine_network}
\end{figure}

The Florentine families network (Figure~\ref{fig:Florentine_network}) exhibits several structural features that are relevant for latent space models. Although relatively small, it contains pronounced heterogeneity in node degree, cohesive groups, and highly central families that acted as brokers between different social circles. Among these, the Medici family occupies a dominant position, serving as a historical example of a node with exceptional structural influence.

This network therefore provides a natural setting for evaluating the proposed model. In particular, we investigate whether the introduction of nodal metric effects allows the model to distinguish structurally dominant families through the estimated values of $\xi^{(i)}$, while simultaneously preserving an accurate latent geometric representation of the network. We compare the proposed formulation with the classical latent space model across different latent geometries, assessing both predictive performance and the ability to reproduce structural properties of the observed network.

\begin{figure}[H]
\centering
\begin{minipage}{0.435\textwidth}
\centering
\includegraphics[width=\linewidth]{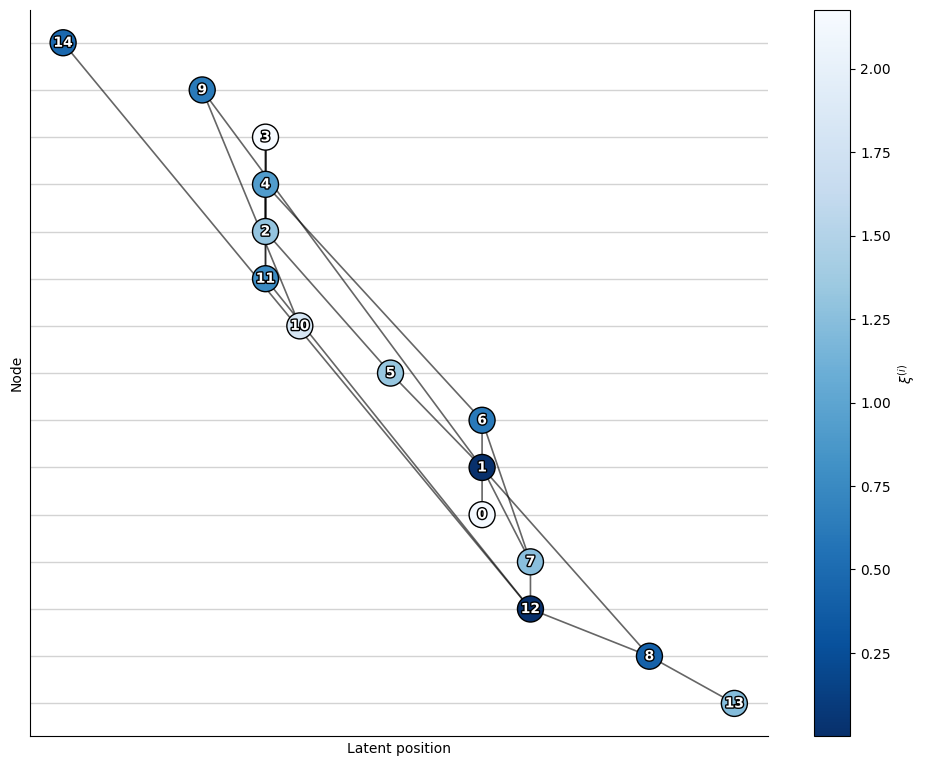}
\captionof*{figure}{(a) Latent space $\mathbb{R}^{1}$. }
\end{minipage}
\begin{minipage}{0.435\textwidth}
\centering
\includegraphics[width=\linewidth]{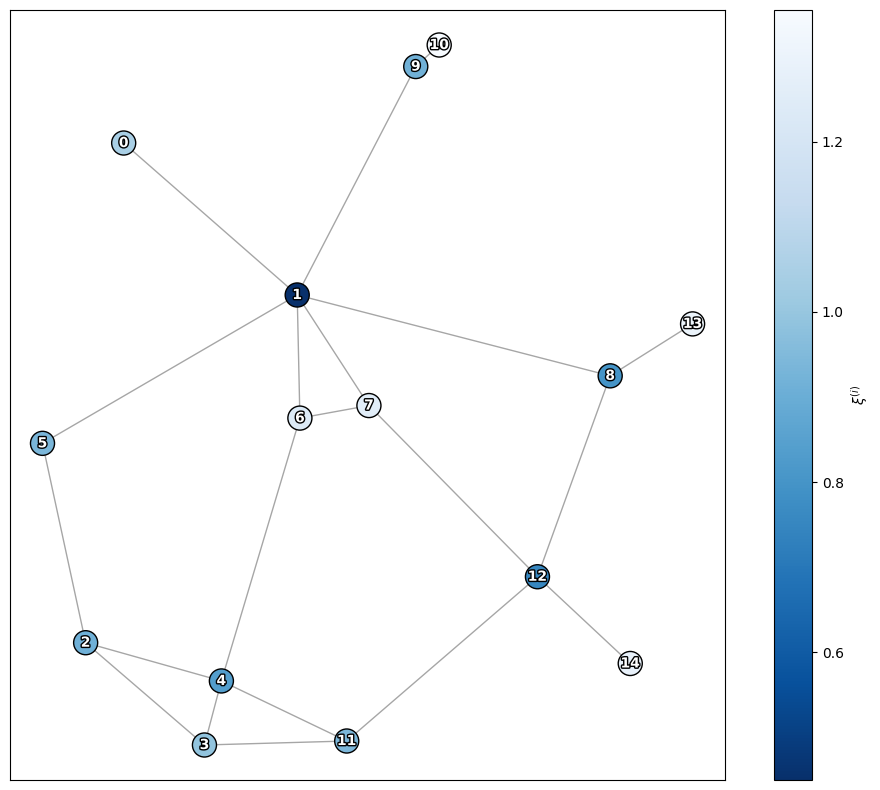}
\captionof*{figure}{(b) Latent space $\mathbb{R}^{2}$.}
\end{minipage}
\hfill
\begin{minipage}{0.435\textwidth}
\centering
\includegraphics[width=\linewidth]{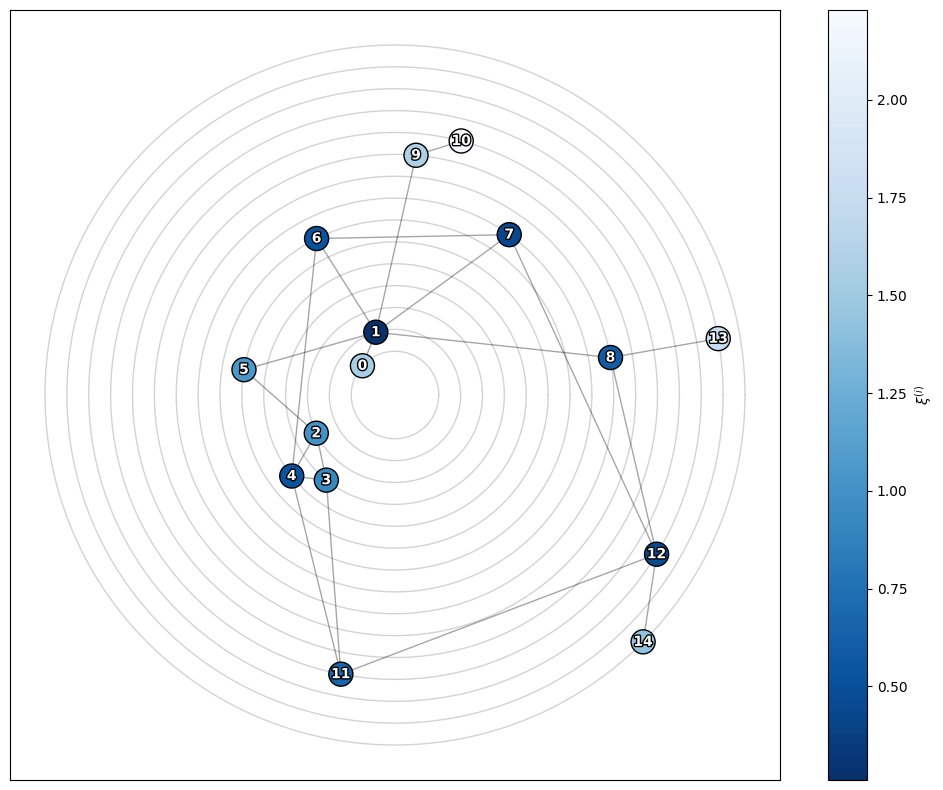}
\captionof*{figure}{(c) Latent space $\mathbb{S}^{1}$.}
\end{minipage}
\begin{minipage}{0.435\textwidth}
\centering
\includegraphics[width=\linewidth]{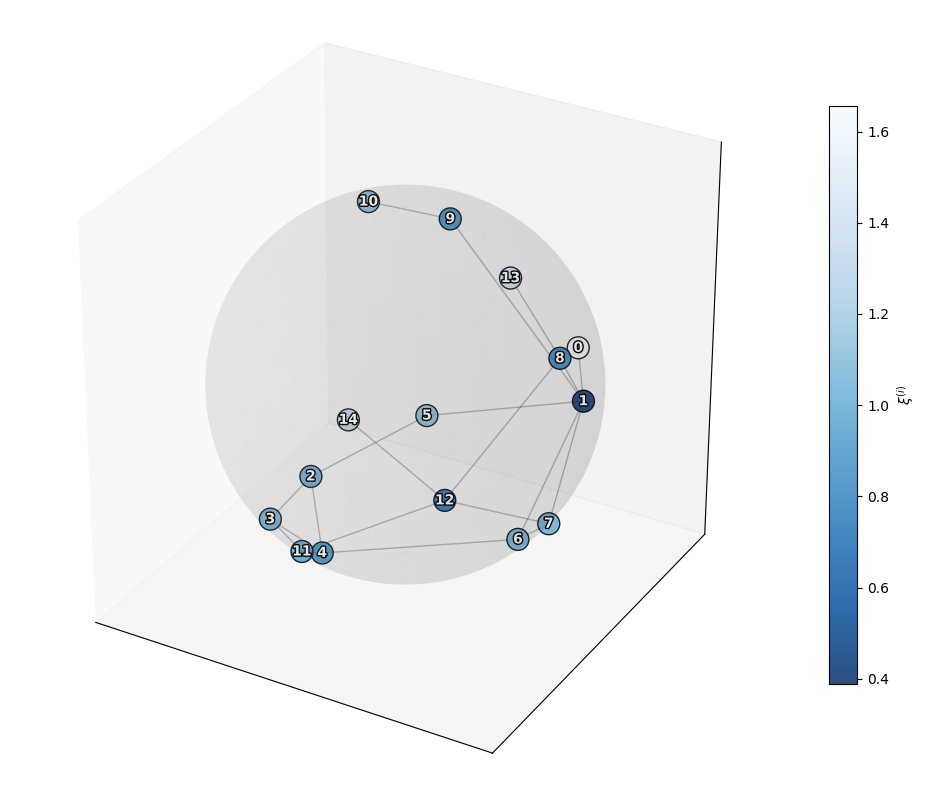}
\captionof*{figure}{(d) Latent space $\mathbb{S}^{2}$.}
\end{minipage}
\hfill
\begin{minipage}{0.435\textwidth}
\centering
\includegraphics[width=\linewidth]{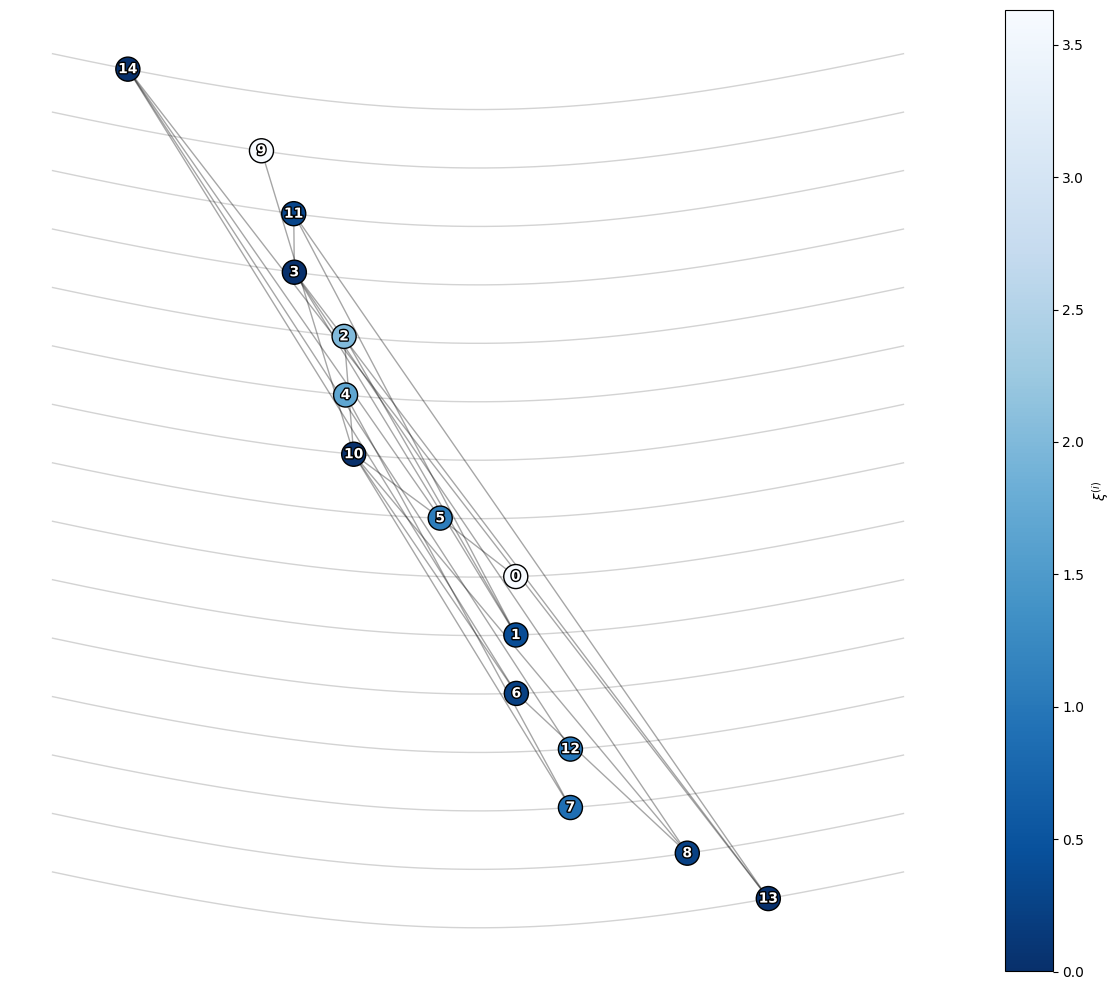}
\captionof*{figure}{(e) Latent space $\mathbb{H}^{1}$.}
\end{minipage}
\begin{minipage}{0.435\textwidth}
\centering
\includegraphics[width=\linewidth]{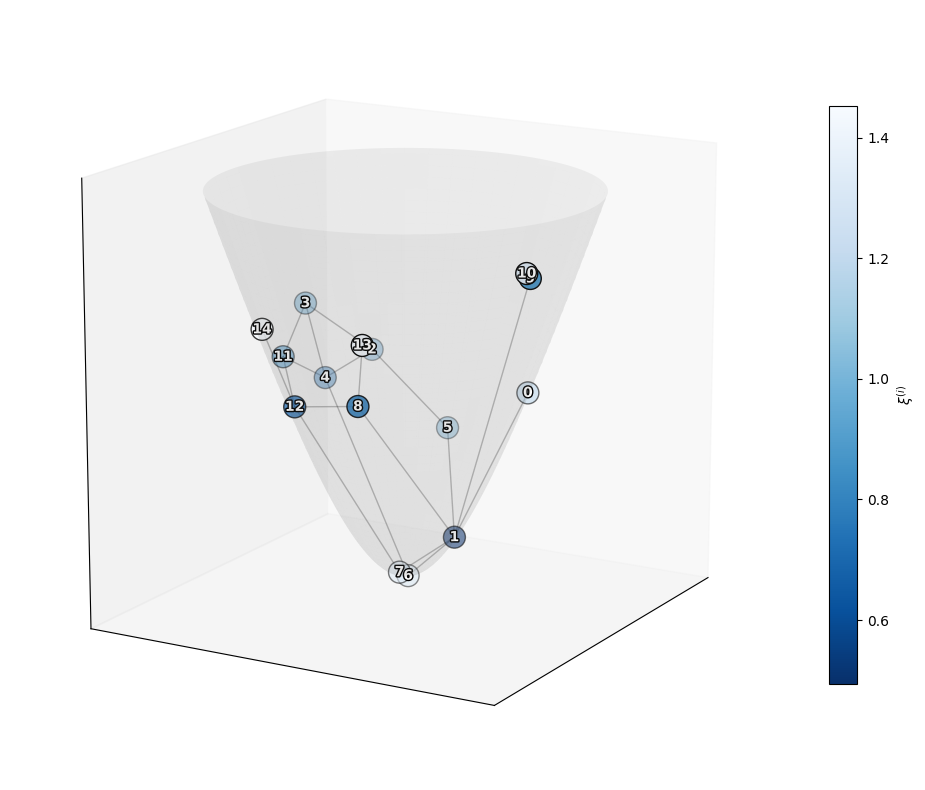}
\captionof*{figure}{(f) Latent space $\mathbb{H}^{2}$.}
\end{minipage}
\hfill
\caption{Visualization of latent spaces for the Florentine families network.}
\label{fig:latentspaces_florentine}
\end{figure}

To evaluate the proposed methodology, we estimate a total of twelve latent space models obtained by combining three latent geometries (Euclidean, spherical, and hyperbolic), two latent dimensions ($d=1$ and $d=2$), and the presence or absence of nodal multiplicative effects. Thus, for each geometry and dimension, we fit both the classical latent space model and the proposed extension with nodal metric effects.

The estimated latent configurations are summarized in Figure~\ref{fig:latentspaces_florentine}. For clarity, only six latent space representations are displayed, corresponding to the combinations of geometry and latent dimension. This simplification is possible because the estimated latent positions under the classical model and the proposed model are the same. Consequently, each panel shows a single latent configuration, while the color of each node represents the estimated value of the corresponding nodal parameter $\xi^{(i)}$. 

Parameter estimation was performed using Algorithms~\ref{alg:classical} and~\ref{alg:plugin}. For all twelve models, we employed $K=20$ random initializations, a maximum of $T=25,000$ optimization iterations, and a learning rate of $\eta=0.01$. The prior specification was fixed as $\alpha_0\sim\mathrm{Normal}(0,1)$ and $\frac{1}{n}\boldsymbol{\xi}\sim\mathrm{Dirichlet}(\mathbf{1}_n)$, while a common latent space diameter constraint of $D=10.0$ was imposed across all geometries to ensure a fair comparison between models.

\begin{figure}[H]
\centering
\includegraphics[width=0.995\linewidth]{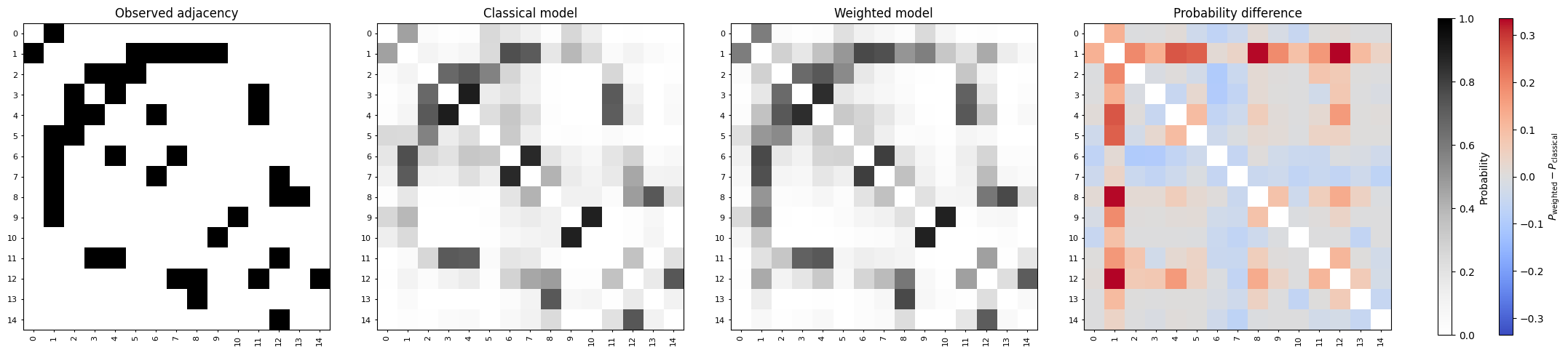}
\caption{Observed and fitted adjacency matrices for the Florentine families network.}
\label{fig:Florentine_ProbMatrix_R2}
\end{figure}

Figure~\ref{fig:Florentine_ProbMatrix_R2} compares the observed adjacency matrix with the fitted probability matrices obtained under the classical and the proposed latent space models in the Euclidean geometry with $d=2$. Overall, both models recover the main connectivity patterns of the observed network, assigning high connection probabilities to the observed edges while maintaining low probabilities for absent links. Nevertheless, the introduction of nodal multiplicative effects produces localized changes in the fitted probability matrix that cannot be explained solely by the latent positions. For example, node~1 exhibits systematically higher connection probabilities under the proposed model, reflecting its relatively small estimated value of $\xi^{(1)}$, which contracts its effective latent distances to the remaining nodes. In contrast, node~6 receives a comparatively large value of $\xi^{(6)}$, expanding its effective distances and consequently reducing its connection probabilities, despite occupying a relatively central position in the latent space $\mathbb{R}^{2}$. 
\begin{figure}[H]
\centering
\includegraphics[width=0.8\linewidth]{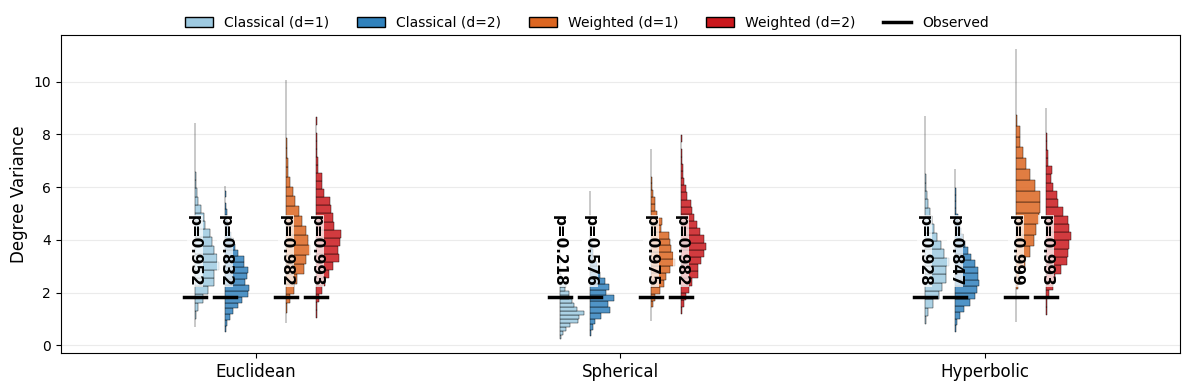}
\caption{Comparison of degree variance for the fitted latent space models on the Florentine families network.}
\label{fig:Florentine_PPC_degreevariance}
\end{figure}

Figure~\ref{fig:Florentine_PPC_degreevariance} presents a predictive check based on the degree variance for the twelve fitted models. Among the considered specifications, the weighted models consistently reproduce the observed degree variance more accurately than their classical counterparts, particularly in one-dimensional spherical geometry, where the observed statistic lies close to the center of the predictive distribution. This improvement reflects the ability of the nodal effects to capture the heterogeneity in node connectivity that is difficult to explain through latent positions alone. We do not present additional posterior predictive checks for other network statistics because the estimation procedure adopted in this work is based on point estimation rather than full Bayesian posterior inference. Consequently, the predictive distributions are generated around a single parameter estimate and do not adequately account for parameter uncertainty, which frequently leads to systematic underestimation or overestimation of the variability of network statistics. 

\begin{figure}[!htb]
\centering
\includegraphics[width=0.90\linewidth]{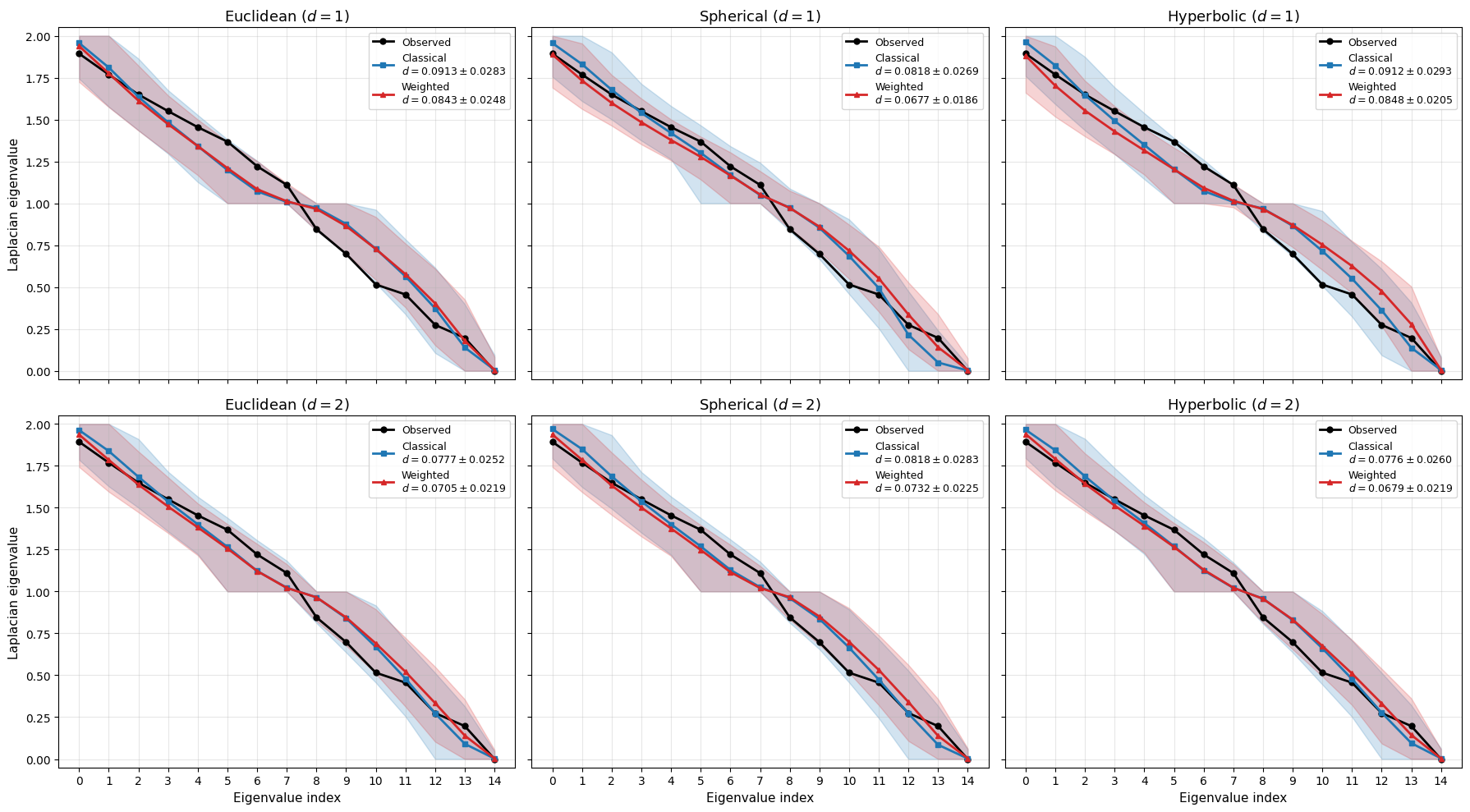}
\caption{Evaluation of the Laplacian spectrum in the Florentine families network.}
\label{fig:florentine_laplacian_spectrum}
\end{figure}

Figure~\ref{fig:florentine_laplacian_spectrum} compares the normalized Laplacian spectrum of the observed network with the spectra generated from the fitted latent space models. As a quantitative measure of structural similarity, we define the normalized Laplacian spectral distance as:
\begin{equation*}
d(\lambda,\hat{\lambda}) := \frac{|\lambda-\hat{\lambda}|_2}{2\sqrt{n}},
\end{equation*}
where $\lambda$ and $\hat{\lambda}$ denote the ordered eigenvalues of the normalized Laplacian matrices of the observed and simulated networks, respectively. Since every eigenvalue of the normalized Laplacian belongs to the interval $[0,2]$, the maximum possible Euclidean distance between two spectra is $2\sqrt{n}$, implying that the proposed distance is naturally bounded in $[0,1]$. In Figure~\ref{fig:florentine_laplacian_spectrum}, the black curve corresponds to the observed spectrum, the solid colored curves represent the average spectrum over 2500 simulated networks generated from each fitted model, and the shaded bands indicate the range of simulated spectra. The legends additionally report the mean Laplacian distance together with its standard deviation. Across all geometries and latent dimensions, the proposed model consistently produces smaller spectral distances than the corresponding classical latent space model, indicating a better reproduction of the global structural properties of the network. The best overall performance is achieved by the spherical model with one latent dimension and nodal multiplicative effects, which attains the smallest average spectral distance among all twelve fitted models.

Table~\ref{tab:florentine_summary} summarizes the quantitative comparison of the twelve fitted models. The information criterion is defined as:
\begin{equation*}
\text{IC}=-2\log p(\hat{\Theta}\mid \mathbf{Y})+k\log\left(\frac{n(n-1)}{2}\right), 
\end{equation*}
where $k=1+dn$ for the classical latent space model and $k=n+dn$ for the proposed model, corresponding to the number of estimated free parameters. This criterion balances model fit against model complexity, with smaller values indicating a more parsimonious model. The column AUC ROC (int) reports the area under the ROC curve computed using the complete fitted network, whereas AUC ROC (ext) reports the mean and standard deviation obtained from a five-fold cross-validation procedure in which the model is re-estimated in every fold. Finally, the last column reports the proposed normalized Laplacian spectral distance, averaged over 2500 networks simulated from each fitted model together with its standard deviation. We propose this quantity as an additional criterion for validating models, since it directly evaluates their ability to reproduce the global spectral structure of the observed network rather than only edge-level prediction. The different evaluation metrics highlight complementary aspects of model performance. The spherical one-dimensional classical model achieves the smallest information criterion owing to its lower complexity, whereas the spherical two-dimensional weighted model provides the highest predictive performance, attaining the best internal and external AUC values. In contrast, the proposed weighted spherical model with one latent dimension yields the smallest Laplacian spectral distance, suggesting that it most faithfully reproduces the global topology of the observed network. More generally, the introduction of nodal multiplicative effects systematically improves both predictive accuracy and structural reconstruction, although this gain comes at the expense of increased model complexity reflected in the information criterion.

\begin{table}[!htb]
\centering
\small{\begin{tabular}{ccc|c|c|rr|rr|}
\cline{4-9}
\textbf{} & \textbf{} & \textbf{}& \textbf{IC}& \textbf{\begin{tabular}[c]{@{}c@{}}AUC\\ ROC \\ (int)\end{tabular}} & \multicolumn{2}{c|}{\textbf{\begin{tabular}[c]{@{}c@{}}AUC \\ ROC \\ (ext)\end{tabular}}} & \multicolumn{2}{c|}{\textbf{\begin{tabular}[c]{@{}c@{}}Laplacian \\ distance\end{tabular}}} \\ \hline
\multicolumn{1}{|c|}{\multirow{4}{*}{$\mathbb{R}^{d}$}} & \multicolumn{1}{c|}{\multirow{2}{*}{1}} & $\xcancel{\boldsymbol{\xi}}$ & 160.989& 0.788 & \multicolumn{1}{r|}{0.757} & 0.119& \multicolumn{1}{r|}{0.091}& 0.028 \\ \cline{3-9} 
\multicolumn{1}{|c|}{}& \multicolumn{1}{c|}{} & $\boldsymbol{\xi}$ & 213.402& 0.886 & \multicolumn{1}{r|}{0.838} & 0.070& \multicolumn{1}{r|}{0.084}& 0.024 \\ \cline{2-9} 
\multicolumn{1}{|c|}{}& \multicolumn{1}{c|}{\multirow{2}{*}{2}} & $\xcancel{\boldsymbol{\xi}}$ & 188.738& 0.985 & \multicolumn{1}{r|}{0.826} & 0.134& \multicolumn{1}{r|}{0.077}& 0.025 \\ \cline{3-9} 
\multicolumn{1}{|c|}{}& \multicolumn{1}{c|}{} & $\boldsymbol{\xi}$ & 247.770& 0.994 & \multicolumn{1}{r|}{0.955} & 0.032& \multicolumn{1}{r|}{0.070}& 0.021 \\ \hline
\multicolumn{1}{|c|}{\multirow{4}{*}{$\mathbb{S}^{d}$}} & \multicolumn{1}{c|}{\multirow{2}{*}{1}} & $\xcancel{\boldsymbol{\xi}}$ & {145.237} & 0.885 & \multicolumn{1}{r|}{0.763} & 0.115& \multicolumn{1}{r|}{0.081}& 0.026 \\ \cline{3-9} 
\multicolumn{1}{|c|}{}& \multicolumn{1}{c|}{} & $\boldsymbol{\xi}$ & 186.250& 0.967 & \multicolumn{1}{r|}{0.886} & 0.082& \multicolumn{1}{r|}{{0.067}} & {0.018}\\ \cline{2-9} 
\multicolumn{1}{|c|}{}& \multicolumn{1}{c|}{\multirow{2}{*}{2}} & $\xcancel{\boldsymbol{\xi}}$ & 191.712& 0.983 & \multicolumn{1}{r|}{0.798} & 0.047& \multicolumn{1}{r|}{0.081}& 0.028 \\ \cline{3-9} 
\multicolumn{1}{|c|}{}& \multicolumn{1}{c|}{} & $\boldsymbol{\xi}$ & 245.962& {0.999}& \multicolumn{1}{r|}{{0.981}}& {0.017} & \multicolumn{1}{r|}{0.073}& 0.022 \\ \hline
\multicolumn{1}{|c|}{\multirow{4}{*}{$\mathbb{H}^{d}$}} & \multicolumn{1}{c|}{\multirow{2}{*}{1}} & $\xcancel{\boldsymbol{\xi}}$ & 215.741& 0.383 & \multicolumn{1}{r|}{0.364} & 0.135& \multicolumn{1}{r|}{0.091}& 0.029 \\ \cline{3-9} 
\multicolumn{1}{|c|}{}& \multicolumn{1}{c|}{} & $\boldsymbol{\xi}$ & 240.866& 0.649 & \multicolumn{1}{r|}{0.440} & 0.153& \multicolumn{1}{r|}{0.084}& 0.020 \\ \cline{2-9} 
\multicolumn{1}{|c|}{}& \multicolumn{1}{c|}{\multirow{2}{*}{2}} & $\xcancel{\boldsymbol{\xi}}$ & 194.502& 0.974 & \multicolumn{1}{r|}{0.972} & 0.048& \multicolumn{1}{r|}{0.077}& 0.026 \\ \cline{3-9} 
\multicolumn{1}{|c|}{}& \multicolumn{1}{c|}{} & $\boldsymbol{\xi}$ & 249.434& 0.992 & \multicolumn{1}{r|}{0.956} & 0.039& \multicolumn{1}{r|}{0.067}& 0.021 \\ \hline
\end{tabular}}
\caption{Summary metrics of the twelve models (geometry, dimension, nodal effects) for the Florentine families network. The external AUC-ROC (AUC ROC (ext)) is reported as the mean and standard deviation over 5-fold cross-validation. The Laplacian distance is reported as the mean and standard deviation computed from 2500 simulated networks generated under each fitted~model.}
\label{tab:florentine_summary}
\end{table}

\subsection{Karate}

\begin{figure}[!htb]
\centering
\begin{minipage}{0.445\textwidth}
\centering
\includegraphics[width=\linewidth]{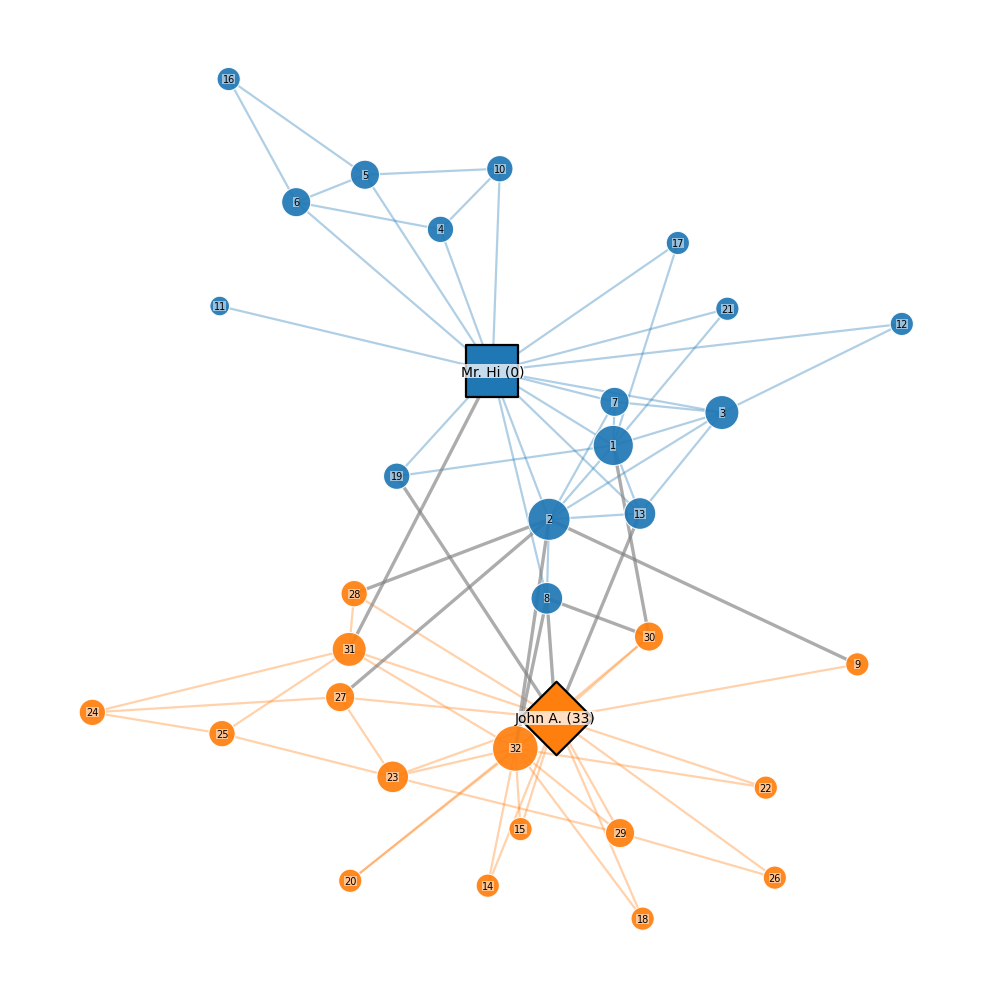}
\captionof*{figure}{(a) Network data. }
\end{minipage} \hfill
\begin{minipage}{0.445\textwidth}
\centering
\includegraphics[width=\linewidth]{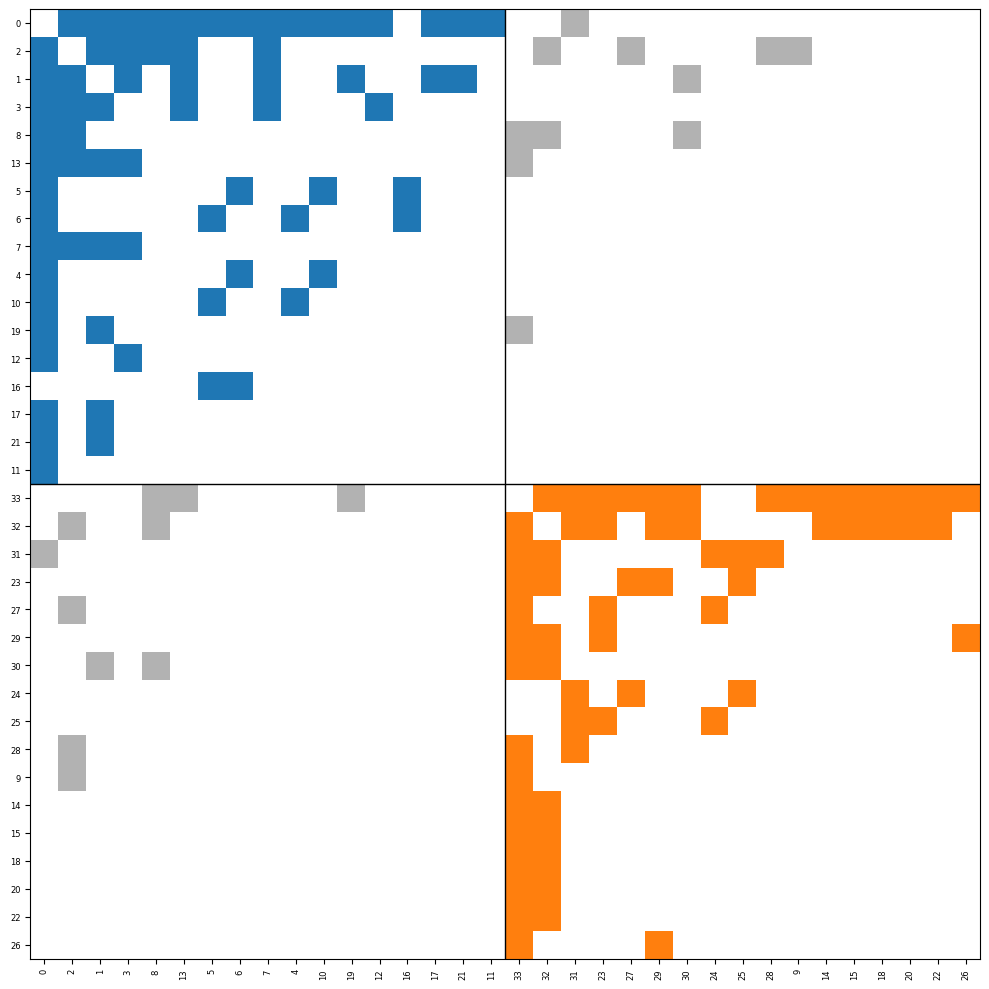}
\captionof*{figure}{(b) Adjacency matrix. }
\end{minipage}
\caption{Visualization of the Zachary karate club network.}
\label{fig:Karate_network}
\end{figure}

The Zachary karate club network is one of the most widely studied benchmark datasets in network science. Originally collected by \cite{zachary1977information}, the network represents the friendship relationships among the 34 members of a university karate club that eventually split into two factions following a conflict between the club instructor (Mr. Hi) and the administrator (John A.). Because the final division of the club is known, this network has become a canonical example for evaluating community detection methods and statistical models for network data. Figure~\ref{fig:Karate_network} presents the network along with its adjacency matrix, separating the two communities by means of colors. As in the Florentine families example, we fit the twelve latent space models obtained by combining the classical and weighted formulations, three latent geometries (Euclidean, spherical, and hyperbolic), and latent dimensions $d=1$ and $d=2$. Besides comparing predictive and structural goodness-of-fit across these models, we further investigate their ability to recover the underlying community structure, exploiting the well-established clustering properties of this network.

\begin{figure}[H]
\centering
\begin{minipage}{0.435\textwidth}
\centering
\includegraphics[width=\linewidth]{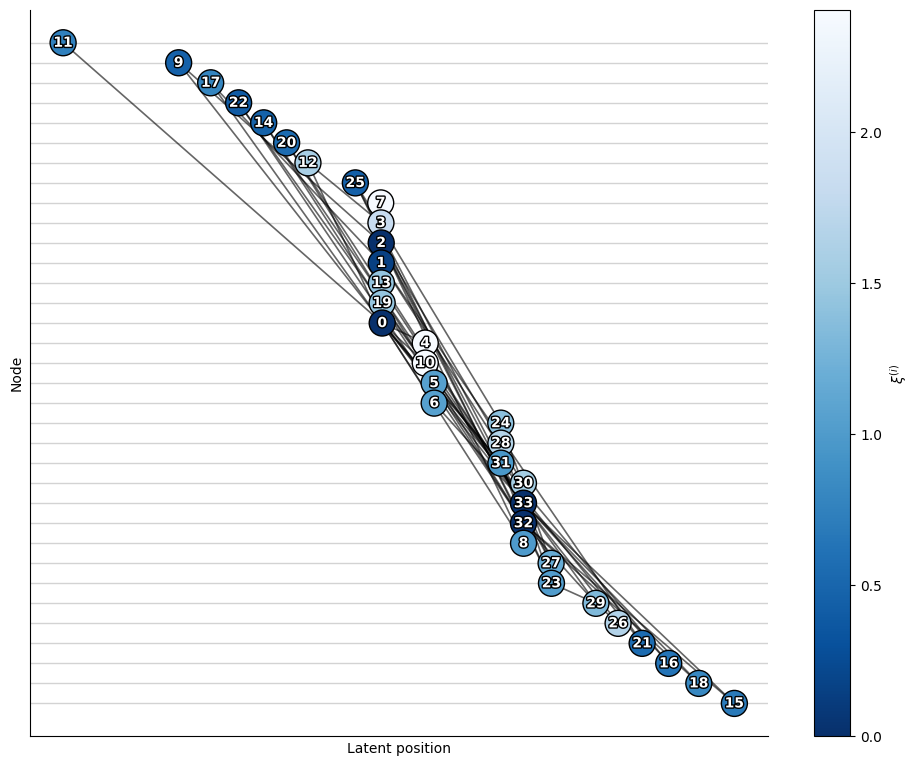}
\captionof*{figure}{(a) Latent space $\mathbb{R}^{1}$. }
\end{minipage}
\begin{minipage}{0.435\textwidth}
\centering
\includegraphics[width=\linewidth]{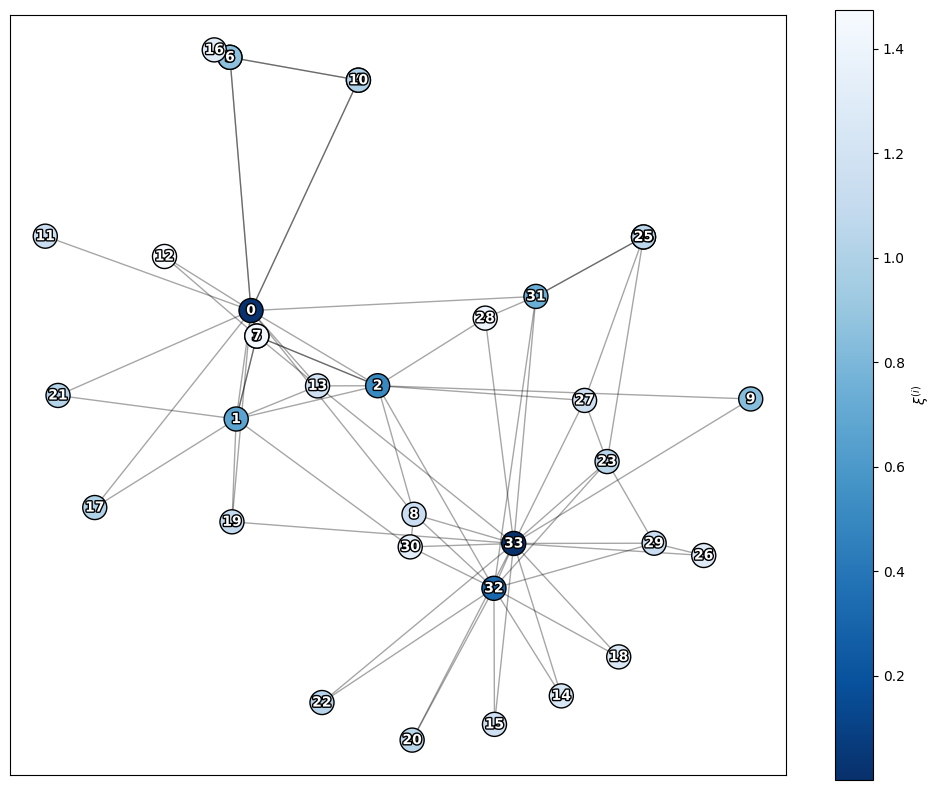}
\captionof*{figure}{(b) Latent space $\mathbb{R}^{2}$.}
\end{minipage}
\hfill
\begin{minipage}{0.435\textwidth}
\centering
\includegraphics[width=\linewidth]{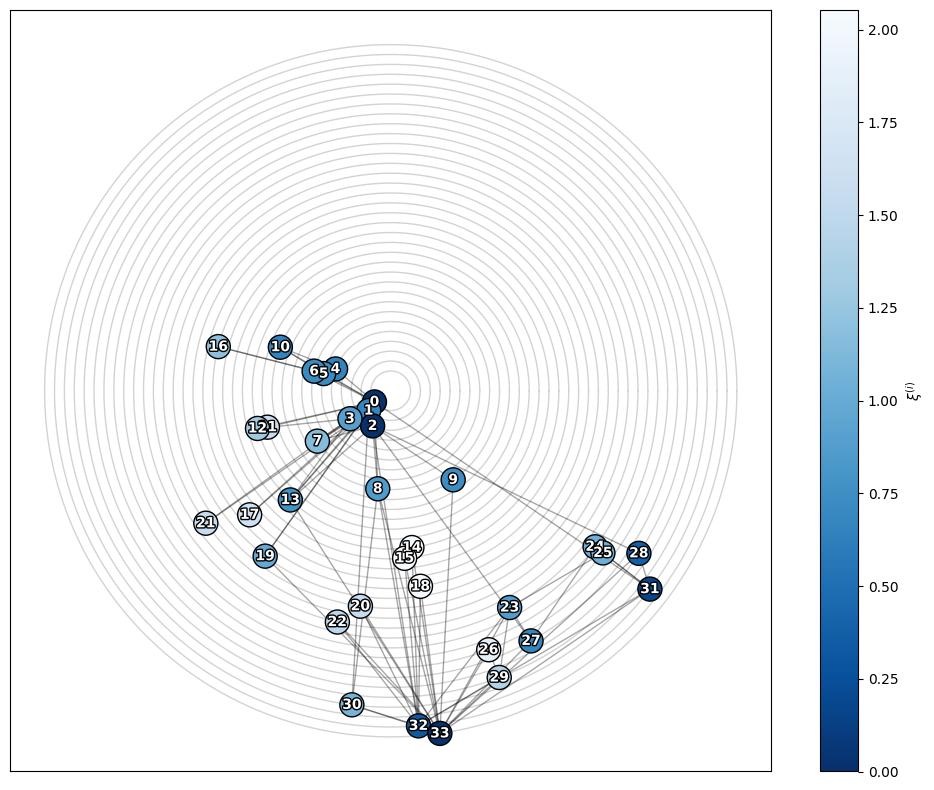}
\captionof*{figure}{(c) Latent space $\mathbb{S}^{1}$.}
\end{minipage}
\begin{minipage}{0.435\textwidth}
\centering
\includegraphics[width=\linewidth]{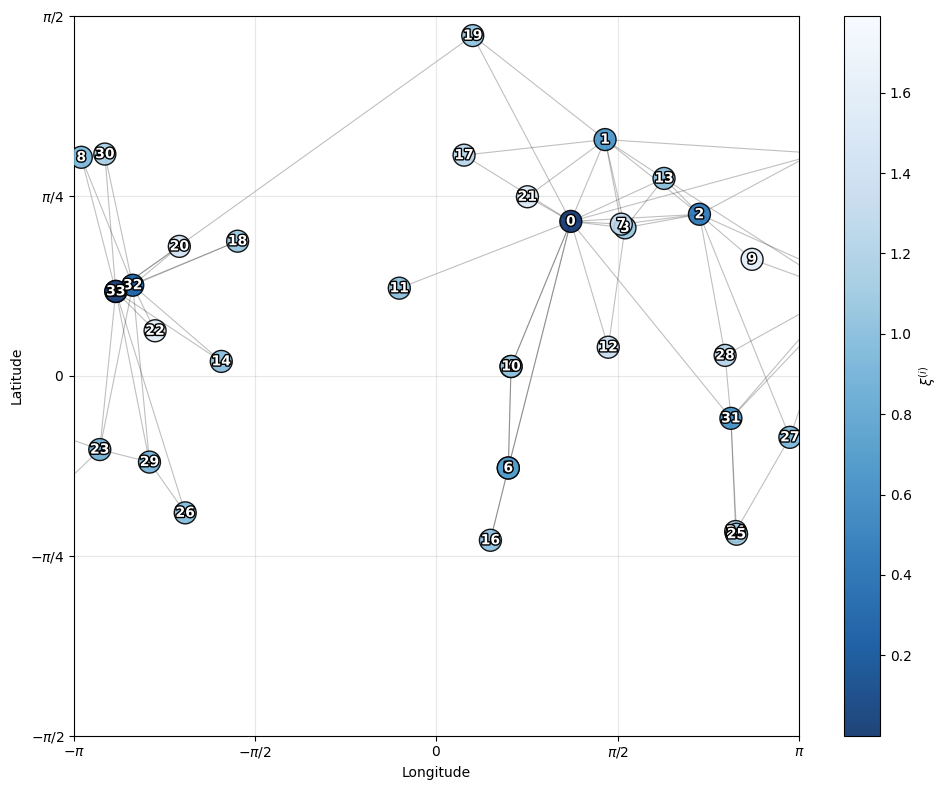}
\captionof*{figure}{(d) Latent space $\mathbb{S}^{2}$.}
\end{minipage}
\hfill
\begin{minipage}{0.435\textwidth}
\centering
\includegraphics[width=\linewidth]{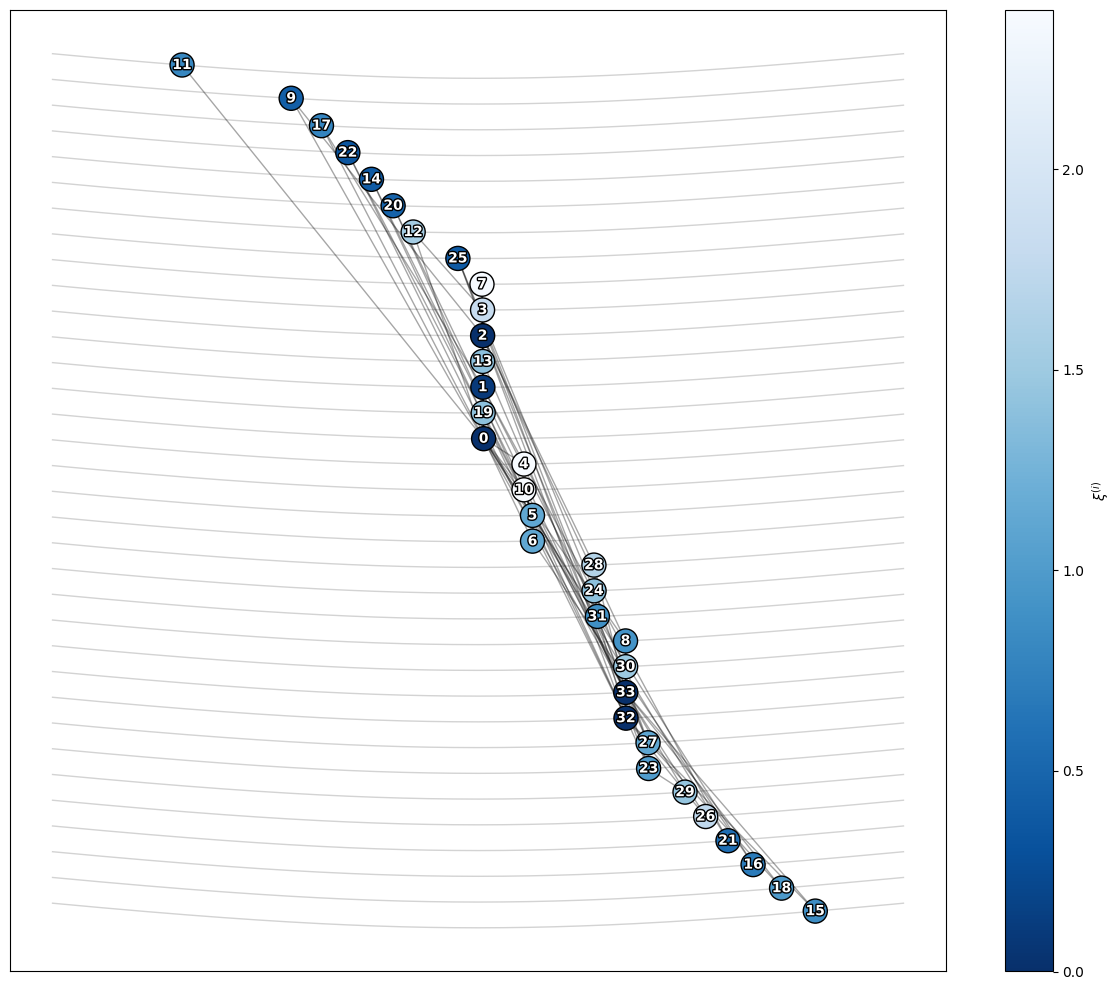}
\captionof*{figure}{(e) Latent space $\mathbb{H}^{1}$.}
\end{minipage}
\begin{minipage}{0.435\textwidth}
\centering
\includegraphics[width=\linewidth]{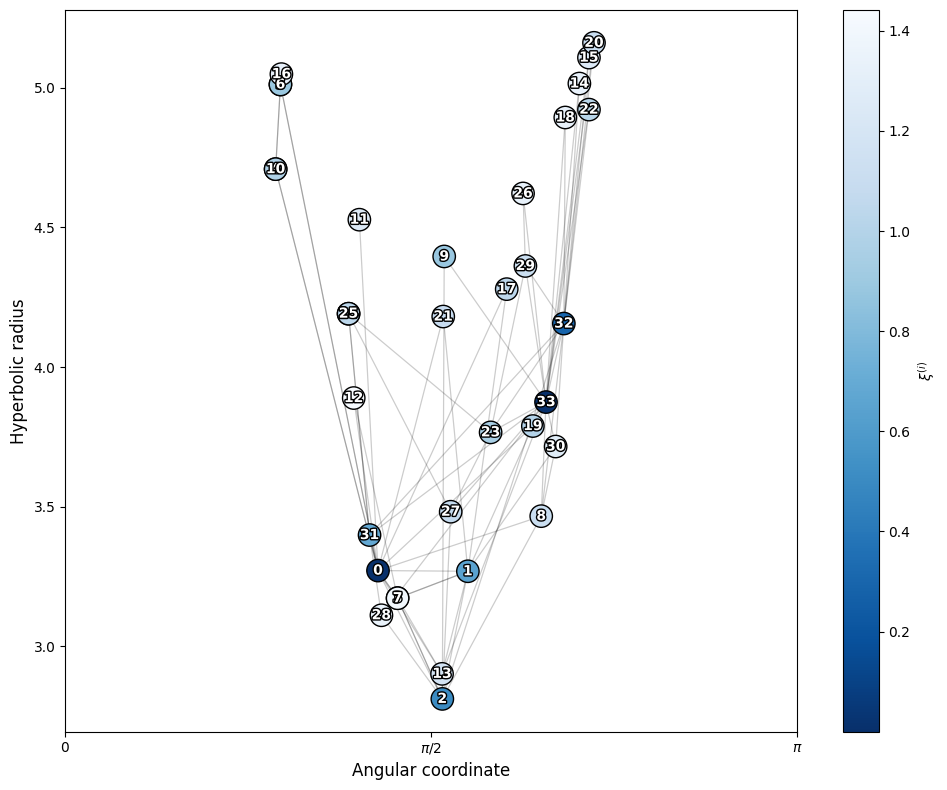}
\captionof*{figure}{(f) Latent space $\mathbb{H}^{2}$.}
\end{minipage}
\hfill
\caption{Visualization of latent spaces for the Karate network.}
\label{fig:latentspaces_karate}
\end{figure}

Figure~\ref{fig:latentspaces_karate} presents the inferred latent spaces for the twelve fitted models. Since the latent embeddings obtained by the classical and weighted formulations are identical, only one latent configuration is displayed for each combination of geometry and latent dimension. In every panel, the color of each node represents the estimated nodal effect~$\xi^{(i)}$, with lighter colors indicating larger values.

\begin{figure}[H]
\centering
\includegraphics[width=0.95\linewidth]{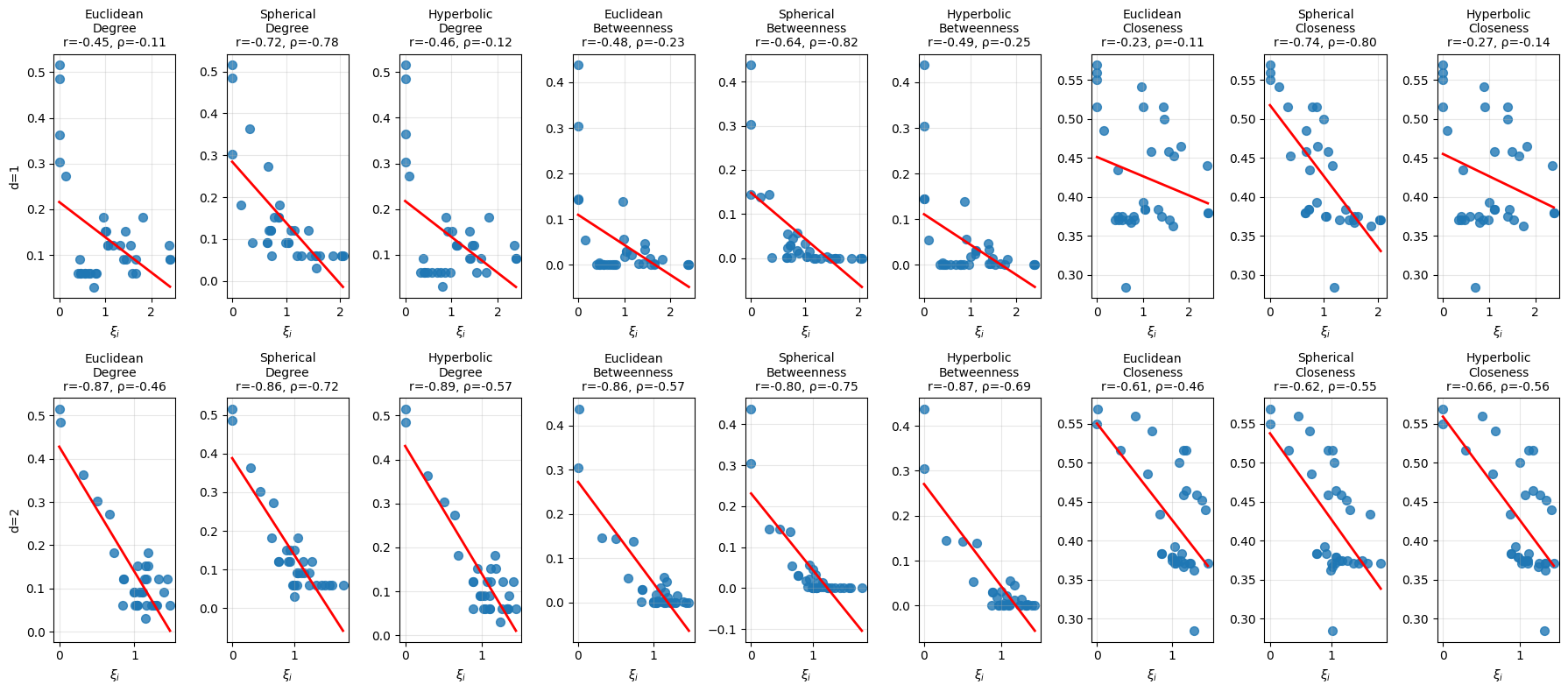}
\caption{Comparison between observed node centrality measures and node multiplicative effects for the fitted latent space models on the Karate network.}
\label{fig:karate_centrality_xi}
\end{figure}

An immediate observation from the inferred latent spaces (Figure~\ref{fig:latentspaces_karate}) is that nodes 0 (Mr. Hi) and 33 (John A.), corresponding to the leaders of the two factions formed after the club split, consistently receive the darkest colors, indicating the smallest estimated values of $\xi^{(i)}$. These two individuals are also known to occupy the most influential and structurally central positions in the network, suggesting that the proposed nodal effects capture an aspect of node importance that is complementary to the latent geometric embedding. This observation motivates a more systematic investigation of the relationship between the estimated nodal effects and classical centrality measures. 

Figure~\ref{fig:karate_centrality_xi} presents, for each of the twelve fitted models, the estimated values of $\xi^{(i)}$ against three standard node centrality measures: degree, betweenness, and closeness. Each panel reports both the Pearson correlation coefficient ($r$), measuring linear association, and the Spearman rank correlation coefficient ($\rho$), measuring monotonic dependence, while the red line corresponds to the least-squares linear regression. Remarkably, the fitted regression exhibits a negative slope in every model and for every centrality measure, revealing a consistent inverse relationship between $\xi^{(i)}$ and node centrality. This behavior is a direct consequence of the proposed model: nodes with smaller values of $\xi^{(i)}$ effectively contract their latent distances to the rest of the network, increasing their connection probabilities and consequently becoming more central, whereas larger values of $\xi^{(i)}$ expand these distances, reducing their structural influence. The association becomes substantially stronger in the two-dimensional models, where both Pearson and Spearman correlations attain consistently larger magnitudes across all geometries, indicating that richer latent representations allow the nodal effects to capture centrality more accurately. 

\begin{figure}[!htb]
\centering
\includegraphics[width=0.85\linewidth]{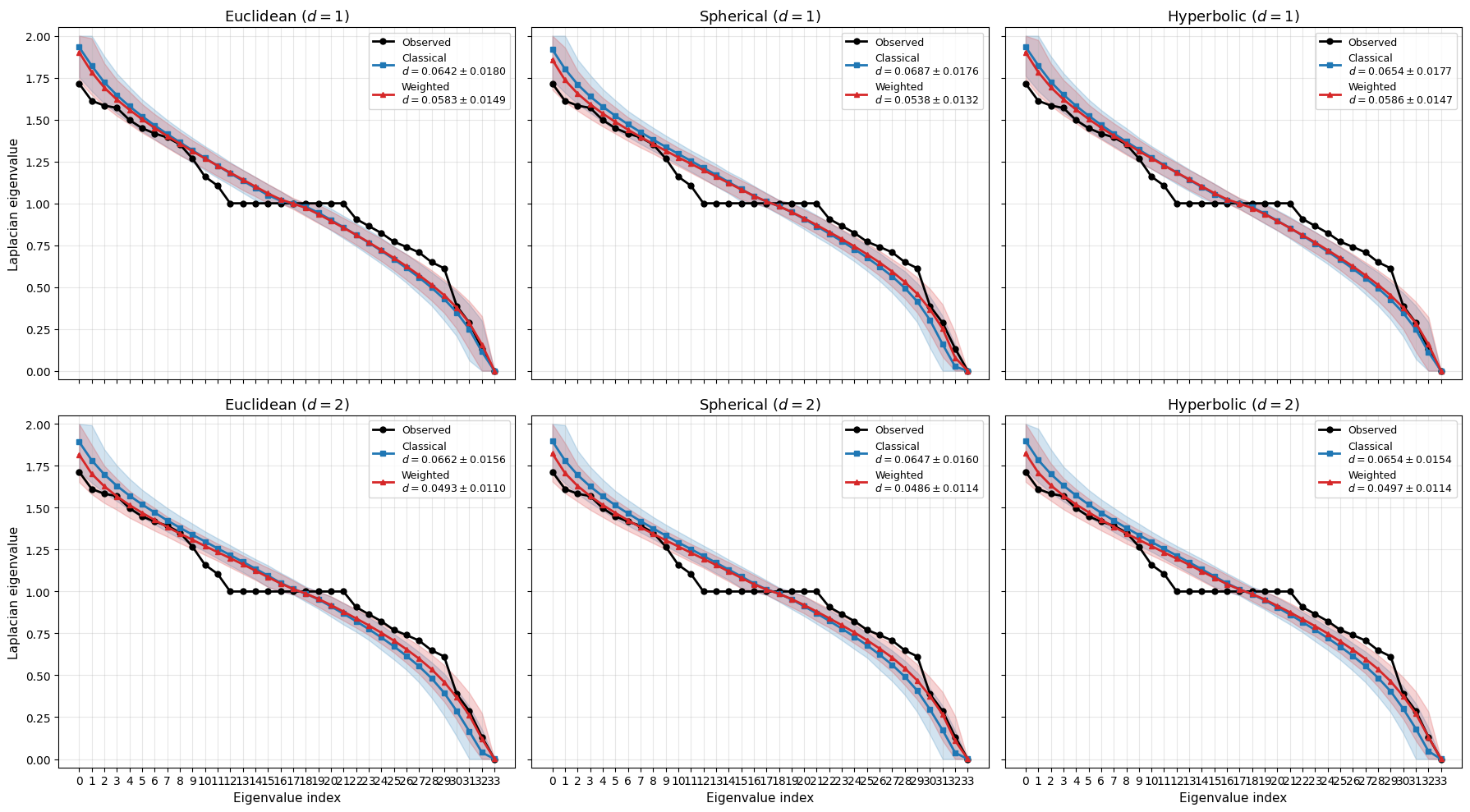}
\caption{Evaluation of the Laplacian spectrum in the Karate network.}
\label{fig:karate_laplacian_spectrum}
\end{figure}

\begin{table}[!b]
\centering
\small{\begin{tabular}{ccc|c|r|rr|rr|}
\cline{4-9}
\textbf{} & \textbf{} & \textbf{}& \textbf{IC} & \multicolumn{1}{c|}{\textbf{\begin{tabular}[c]{@{}c@{}}AUC\\ ROC \\ (int)\end{tabular}}} & \multicolumn{2}{c|}{\textbf{\begin{tabular}[c]{@{}c@{}}AUC \\ ROC \\ (ext)\end{tabular}}} & \multicolumn{2}{c|}{\textbf{\begin{tabular}[c]{@{}c@{}}Laplacian \\ distance\end{tabular}}} \\ \hline
\multicolumn{1}{|c|}{\multirow{4}{*}{$\mathbb{R}^{d}$}} & \multicolumn{1}{c|}{\multirow{2}{*}{1}} & $\xcancel{\boldsymbol{\xi}}$ & 627.204 & 0.736& \multicolumn{1}{r|}{0.639} & 0.079& \multicolumn{1}{r|}{0.064}& 0.018 \\ \cline{3-9} 
\multicolumn{1}{|c|}{}& \multicolumn{1}{c|}{} & $\boldsymbol{\xi}$ & 787.435 & 0.857& \multicolumn{1}{r|}{0.797} & 0.033& \multicolumn{1}{r|}{0.058}& 0.014 \\ \cline{2-9} 
\multicolumn{1}{|c|}{}& \multicolumn{1}{c|}{\multirow{2}{*}{2}} & $\xcancel{\boldsymbol{\xi}}$ & 700.646 & 0.920& \multicolumn{1}{r|}{0.810} & 0.019& \multicolumn{1}{r|}{0.066}& 0.015 \\ \cline{3-9} 
\multicolumn{1}{|c|}{}& \multicolumn{1}{c|}{} & $\boldsymbol{\xi}$ & 850.169 & 0.985& \multicolumn{1}{r|}{0.965} & 0.015& \multicolumn{1}{r|}{0.048}& 0.011 \\ \hline
\multicolumn{1}{|c|}{\multirow{4}{*}{$\mathbb{S}^{d}$}} & \multicolumn{1}{c|}{\multirow{2}{*}{1}} & $\xcancel{\boldsymbol{\xi}}$ & 551.248 & 0.860& \multicolumn{1}{r|}{0.800} & 0.032& \multicolumn{1}{r|}{0.068}& 0.017 \\ \cline{3-9} 
\multicolumn{1}{|c|}{}& \multicolumn{1}{c|}{} & $\boldsymbol{\xi}$ & 691.963 & 0.946& \multicolumn{1}{r|}{0.921} & 0.021& \multicolumn{1}{r|}{0.053}& 0.013 \\ \cline{2-9} 
\multicolumn{1}{|c|}{}& \multicolumn{1}{c|}{\multirow{2}{*}{2}} & $\xcancel{\boldsymbol{\xi}}$ & 703.648 & 0.918& \multicolumn{1}{r|}{0.823} & 0.031& \multicolumn{1}{r|}{0.064}& 0.016 \\ \cline{3-9} 
\multicolumn{1}{|c|}{}& \multicolumn{1}{c|}{} & $\boldsymbol{\xi}$ & 840.949 & 0.984& \multicolumn{1}{r|}{0.965} & 0.016& \multicolumn{1}{r|}{0.048}& 0.011 \\ \hline
\multicolumn{1}{|c|}{\multirow{4}{*}{$\mathbb{H}^{d}$}} & \multicolumn{1}{c|}{\multirow{2}{*}{1}} & $\xcancel{\boldsymbol{\xi}}$ & 628.738 & 0.735& \multicolumn{1}{r|}{0.734} & 0.051& \multicolumn{1}{r|}{0.065}& 0.017 \\ \cline{3-9} 
\multicolumn{1}{|c|}{}& \multicolumn{1}{c|}{} & $\boldsymbol{\xi}$ & 790.147 & 0.856& \multicolumn{1}{r|}{0.796} & 0.028& \multicolumn{1}{r|}{0.058}& 0.014 \\ \cline{2-9} 
\multicolumn{1}{|c|}{}& \multicolumn{1}{c|}{\multirow{2}{*}{2}} & $\xcancel{\boldsymbol{\xi}}$ & 706.666 & 0.920& \multicolumn{1}{r|}{0.919} & 0.017& \multicolumn{1}{r|}{0.065}& 0.015 \\ \cline{3-9} 
\multicolumn{1}{|c|}{}& \multicolumn{1}{c|}{} & $\boldsymbol{\xi}$ & 861.441 & 0.976& \multicolumn{1}{r|}{0.950} & 0.015& \multicolumn{1}{r|}{0.049}& 0.011 \\ \hline
\end{tabular}}
\caption{Summary metrics of the twelve models (geometry, dimension, nodal effects) for the Karate network. The external AUC-ROC (AUC ROC (ext)) is reported as the mean and standard deviation over 5-fold cross-validation. The Laplacian distance is reported as the mean and standard deviation computed from 2500 simulated networks generated under each fitted model.}
\label{tab:karate_summary}
\end{table}

The overall model comparison is summarized in Figure~\ref{fig:karate_laplacian_spectrum} and Table~\ref{tab:karate_summary}. As observed for the Florentine families network, the weighted latent space models consistently produce a better approximation of the normalized Laplacian spectrum than their classical counterparts, yielding smaller average spectral distances across all geometries and latent dimensions. In particular, the two-dimensional weighted Euclidean and spherical models achieve the best structural reconstruction, both attaining an average Laplacian distance of approximately $0.048\pm0.011$, while the corresponding classical models remain above $0.064$. This improvement is also reflected in the predictive performance, where the weighted models substantially increase both the in-sample and cross-validated AUC values. For example, the Euclidean model with $d=2$ improves the external AUC from $0.810\pm0.019$ to $0.965\pm0.015$ after incorporating nodal multiplicative effects. In contrast, the information criterion systematically favors the simpler classical models due to the additional parameters introduced by the proposed formulation, with the one-dimensional spherical model achieving the lowest IC. These results highlight the trade-off between model complexity and goodness of fit: although the weighted models incur a larger complexity penalty, they consistently provide a more faithful reconstruction of both the edge structure and the global topology of the network, as quantified by the proposed normalized Laplacian spectral~distance.

The Karate network provides a natural benchmark for evaluating community detection algorithms, since its true partition into two factions is historically known. To assess whether the inferred probability matrices preserve this community structure, we performed spectral clustering on the estimated probability matrix $\widehat{\mathbf{P}}$ obtained from each fitted model. Specifically, after symmetrizing the estimated probability matrix and setting the diagonal entries to one, we applied the spectral clustering algorithm proposed by \cite{ng2002spectral}, using the inferred probabilities as a precomputed affinity matrix and fixing the number of clusters to two. The resulting partition was then compared with the true club division using four complementary clustering metrics: the Adjusted Rand Index (ARI, \cite{hubert1985comparing}), the Normalized Mutual Information (NMI, \cite{strehl2002cluster}), the Fowlkes--Mallows Index (FMI, \cite{fowlkes1983method}), and the clustering accuracy after optimal label matching. These measures evaluate different aspects of clustering agreement, ranging from pairwise consistency (ARI and FMI) to information-theoretic similarity (NMI) and classification performance (Accuracy), thereby providing a comprehensive assessment of the recovered community structure.

The clustering results are summarized in Table~\ref{tab:karate_centrality}. In contrast to the predictive and structural validation metrics discussed previously, the proposed weighted models do not consistently improve the recovery of the two communities. In fact, the best clustering performance is generally achieved by the classical two-dimensional models, particularly in the Euclidean and hyperbolic geometries, which attain $\mathrm{ARI}=0.771$, $\mathrm{NMI}=0.732$, $\mathrm{FMI}=0.883$, and an accuracy of $94.1\%$. The corresponding weighted models exhibit a slight decrease in all clustering metrics, with the Euclidean and hyperbolic two-dimensional models dropping to $\mathrm{ARI}=0.668$ and an accuracy of $91.1\%$. Interestingly, the spherical geometry is essentially unaffected by the inclusion of nodal multiplicative effects, yielding identical clustering results for both formulations. These observations suggest that the additional flexibility introduced by the nodal effects is primarily devoted to modeling heterogeneous node propensities rather than strengthening the separation between communities. Consequently, while the proposed model provides superior edge prediction and a more faithful reconstruction of the global network topology, it does not necessarily enhance community detection when spectral clustering is applied to the inferred probability matrices.

\begin{table}[!htb]
\centering
\small{\begin{tabular}{ccc|c|c|c|c|}
\cline{4-7}
\textbf{} & \textbf{} & \textbf{}& \textbf{ARI} & \textbf{NMI} & \textbf{FMI} & \textbf{Accuracy} \\ \hline
\multicolumn{1}{|c|}{\multirow{4}{*}{$\mathbb{R}^{d}$}} & \multicolumn{1}{c|}{\multirow{2}{*}{1}} & $\xcancel{\boldsymbol{\xi}}$ & 0.258& 0.217& 0.620& 0.764 \\ \cline{3-7} 
\multicolumn{1}{|c|}{}& \multicolumn{1}{c|}{} & $\boldsymbol{\xi}$ & 0.258& 0.217& 0.620& 0.764 \\ \cline{2-7} 
\multicolumn{1}{|c|}{}& \multicolumn{1}{c|}{\multirow{2}{*}{2}} & $\xcancel{\boldsymbol{\xi}}$ & 0.771& 0.732& 0.883& 0.941 \\ \cline{3-7} 
\multicolumn{1}{|c|}{}& \multicolumn{1}{c|}{} & $\boldsymbol{\xi}$ & 0.668& 0.648& 0.831& 0.911 \\ \hline
\multicolumn{1}{|c|}{\multirow{4}{*}{$\mathbb{S}^{d}$}} & \multicolumn{1}{c|}{\multirow{2}{*}{1}} & $\xcancel{\boldsymbol{\xi}}$ & 0.771& 0.732& 0.883& 0.941 \\ \cline{3-7} 
\multicolumn{1}{|c|}{}& \multicolumn{1}{c|}{} & $\boldsymbol{\xi}$ & 0.771& 0.732& 0.883& 0.941 \\ \cline{2-7} 
\multicolumn{1}{|c|}{}& \multicolumn{1}{c|}{\multirow{2}{*}{2}} & $\xcancel{\boldsymbol{\xi}}$ & 0.771& 0.732& 0.883& 0.941 \\ \cline{3-7} 
\multicolumn{1}{|c|}{}& \multicolumn{1}{c|}{} & $\boldsymbol{\xi}$ & 0.771& 0.732& 0.883& 0.941 \\ \hline
\multicolumn{1}{|c|}{\multirow{4}{*}{$\mathbb{H}^{d}$}} & \multicolumn{1}{c|}{\multirow{2}{*}{1}} & $\xcancel{\boldsymbol{\xi}}$ & 0.258& 0.217& 0.620& 0.764 \\ \cline{3-7} 
\multicolumn{1}{|c|}{}& \multicolumn{1}{c|}{} & $\boldsymbol{\xi}$ & 0.258& 0.217& 0.620& 0.764 \\ \cline{2-7} 
\multicolumn{1}{|c|}{}& \multicolumn{1}{c|}{\multirow{2}{*}{2}} & $\xcancel{\boldsymbol{\xi}}$ & 0.771& 0.732& 0.883& 0.941 \\ \cline{3-7} 
\multicolumn{1}{|c|}{}& \multicolumn{1}{c|}{} & $\boldsymbol{\xi}$ & 0.668& 0.648& 0.831& 0.911 \\ \hline
\end{tabular}}
\caption{Clustering metrics of the twelve models (geometry, dimension, nodal effects) for the Karate network.}
\label{tab:karate_centrality}
\end{table}

\section{Discussion and conclusions}\label{sec6}

We have introduced a geometric extension of latent space models in which each node is endowed with a latent conformal factor that locally deforms the underlying metric. Rather than assuming that all nodes interact under a common geometry, the proposed framework allows the effective notion of distance to vary across the network while preserving the global Euclidean, spherical, or hyperbolic structure. This construction provides a simple and interpretable mechanism for modeling heterogeneous interaction patterns that cannot be explained solely by latent positions.

Both the theoretical development and the empirical results demonstrate that incorporating nodal metric effects substantially increases the expressive power of latent space models. Across all networks considered in this work, the proposed model consistently achieves better predictive performance than the classical latent space model while simultaneously providing a more faithful reconstruction of the observed Laplacian spectrum. Moreover, the simulation studies show that the deformed metric reproduces node-level structural characteristics, such as degree, betweenness, and closeness centralities, considerably more accurately than the classical formulation. These results suggest that a significant portion of the unexplained variability traditionally attributed to latent positions can instead be captured through local geometric heterogeneity.

An important feature of the proposed approach is that it preserves the probabilistic and geometric interpretation of classical latent space models. The additional parameters do not replace the latent coordinates but rather complement them by introducing a local scaling of distances, allowing the model to distinguish between nodes that occupy similar latent positions but exhibit substantially different levels of connectivity or structural influence. This interpretation naturally links geometric latent models with ideas from differential geometry, where local metric deformations modify distances without changing the underlying topology of the manifold.

The proposed framework also opens several directions for future research. A natural extension consists of incorporating nodal and dyadic covariates into the linear predictor or directly into the metric deformation, allowing observed attributes to influence the local geometry. The methodology can also be generalized to weighted, directed, multilayer, and temporal networks, where metric deformations may evolve across layers or time. From a computational perspective, scalable variational inference algorithms would facilitate applications to substantially larger networks while preserving Bayesian uncertainty quantification.

From a geometric perspective, several interesting generalizations remain unexplored. Instead of restricting attention to fixed Euclidean, spherical, or hyperbolic spaces, one may estimate the curvature directly from the data or consider more general anisotropic (non-conformal) metric deformations, allowing different directions in the latent space to be stretched differently. Likewise, replacing point-valued latent representations by probability distributions would establish a natural connection with information geometry and optimal transport, where similarity between nodes is measured through divergences or Wasserstein distances rather than classical geodesic distances.

Finally, important theoretical questions remain open. These include establishing conditions for global identifiability, posterior consistency, asymptotic normality of the estimators, automatic selection of the latent dimension, and extensions to higher-order network structures such as hypergraphs and simplicial complexes. We believe that these directions provide a promising research agenda toward a broader theory of heterogeneous geometric latent representations for complex networks. 

\section*{Statements and declarations}

The authors declare that they have no financial interests or personal relationships that could influence the work described in this article.

During the preparation of this paper, the authors used ChatGPT-5 to improve the language and readability. After using this tool, the authors reviewed and edited the content as necessary and assume full responsibility for the content of the publication.

\bibliography{references.bib}
\bibliographystyle{apalike}

\end{document}